# Sympathetically cooled molecular ions: from principles to first applications

Bernhard Roth, Stephan Schiller
Institut für Experimentalphysik,
Heinrich–Heine–Universität Düsseldorf,
40225 Düsseldorf, Germany

December 5, 2008

# Contents

# Chapter 1

# Sympathetically cooled molecular ions: from principles to first applications

## 1.1 Introduction

The emerging field of cold molecules offers applications ranging from novel studies of light-molecule interactions (e.g. coherences within rotational states) over interactions between dipolar molecules [1], to ultra-high resolution spectroscopy, and to the study of complex systems [2]. In spectroscopy, a similar qualitative jump is expected as was obtained in atomic spectroscopy after the technique of laser cooling was introduced. Vibrational and rotational levels of suitable molecules have long lifetimes (ms to days) implying potentially huge transition quality factors. Fundamental physics experiments with molecules requiring extreme resolution and accuracy are e.g. tests of quantum electrodynamics (QED), measurement of and search for a space-time variability of the proton-to-electron mass ratio [3], the search for parity violation effects on vibrational transition frequencies [4, 5], and tests of the isotropy of space [6]. Finally, a novel direction is the investigation of collisions of molecules with atoms or other molecules at very low temperatures [7, 8]. This temperature regime represents a unique situation for the study of quantum mechanical details of collisional processes, preferably with the collision partners in well-defined internal states [9].

All of these topics can be studied both with neutral molecules and with molecular ions (Fig. 1.1). Some molecular systems of particular interest, such as the one-electron molecules, are inherently ions (molecular hydrogen ions). At present, the range of molecular ions that can be produced at very low temperature is much wider than for neutral molecules. This situation makes them also interesting for testing methods that may later be extended to neutral molecules as well.

The techniques developed for the laser-cooling of neutral and charged atoms cannot be directly applied to molecules, due to the lack of closed optical transitions. Therefore, new methods for the production of cold molecules are required. Of particular interest are general methods which do not depend on the internal nature of the particles, such as their magnetic or electric moments or their energy level structure, so that they can be applied to a large variety of molecular species, ranging from light diatomics to complex molecules, such as proteins and polymers.

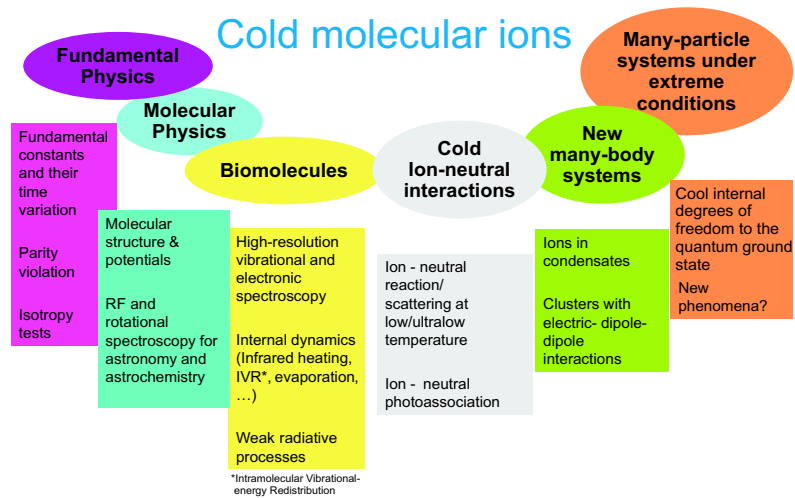

Figure 1.1: Overview of potential applications of cold molecular ions.

Buffer gas (collision) cooling is such a general approach, and cold molecular ions have been produced using cold helium gas already a long time ago. Translational and internal temperatures of a few Kelvin were reached [10]. In other work, buffer gas cooling was combined with multipole linear radiofrequency traps [11]. Here the achieved temperatures are significantly higher, $\sim$15 K. This approach is used by several groups, for studies of spectroscopy and chemical reactions of both small and large molecular ions [12, 13, 14, 15, 16, 17].

## 1.2 Sympathetic cooling

Temperatures significantly lower than those obtained with cryogenic He buffer gas cooling can be achieved by using laser-cooled atomic ions as a charged "buffer-gas". This method is referred to as sympathetic or interaction cooling. Here, two or more different species are trapped simultaneously in an ion trap. One of them is directly cooled by lasers, and the other species will, at least in part, eventually also cool down through long-range Coulomb interactions. Sympathetic cooling was first observed in laser cooling experiments of atomic ions in Penning traps [18, 19] and later in radio frequency traps [20, 21, 22].

One important advantage of sympathetic cooling is that it does not depend on the internal level structure or on the electric or magnetic moment of the particles, but only their mass and charge. Sympathetic cooling of atomic species has been implemented using as atomic coolants laser-cooled $^9$Be$^+$ [23], $^{24}$Mg$^+$ [24, 25, 26], $^{40}$Ca$^+$ [27], $^{114}$Cd$^+$ [28], and $^{138}$Ba$^+$ [29], in both electrostatic and electrodynamic ion traps. All sympathetically cooled species were singly charged, medium-mass atoms, except in one experiment on highly charged xenon [23].

| crystallized state | fluid state |
|---|---|
| BeH$^+$ [34], MgH$^+$, O$_2^+$, MgO$^+$, CaO$^+$ [30, 32], H$_2^+$, HD$^+$, D$_2^+$, H$_3^+$, D$_3^+$, H$_2$D$^+$, HD$_2^+$, BeH$^+$, BeD$^+$, NeH$^+$, NeD$^+$, N$_2^+$, OH$^+$, H$_2$O$^+$, O$_2^+$, HO$_2^+$, ArH$^+$, ArD$^+$, CO$_2^+$, KrH$^+$, KrD$^+$, BaO$^+$, C$_4$F$_8^+$, AF350$^+$, GAH$^+$, R6G$^+$ and fragments [36], [this work] cytochrome c proteins (Cyt$^{12+}$, Cyt$^{17+}$) [37] | NH$_4^+$, H$_2$O$^+$, H$_3$O$^+$, C$_2$H$_5^+$, COH$^+$, O$_2^+$ [29, 33], C$_{60}^+$ [35] |

Table 1.1: Molecular ions sympathetically cooled to date.

Sympathetic ion cooling is equally well applicable to molecular ions. Ground-breaking work has been performed by Drewsen and coworkers, who produced MgH$^+$ molecular ions by reactions of H$_2$ with laser-cooled Mg$^+$, and observed subsequent sympathetic cooling and crystallization [30, 31]. Baba and Waki demonstrated sympathetic cooling of H$_3$O$^+$, NH$_4^+$, O$_2^+$, and C$_2$H$_5^+$ ions, also using Mg$^+$[38]; here the temperatures $T \sim 10\,\text{K}$ corresponded to the liquid state. These two groups also showed the utility of linear quadrupole ion traps for sympathetic cooling.

The second advantage of sympathetic cooling of ions is the notable cooling strength, due to the Coulomb interaction. In our work we showed that using just two atomic ion species, sympathetic cooling can be employed to cool any (singly charged) atomic and molecular ion species in the wide mass range 1 - 470 amu, as well as much heavier ions (up to mass 12,400 amu) if they are highly charged (Table 1.1).

## 1.3 Ion trapping and production of cold molecules

### 1.3.1 Radiofrequency ion traps

Charged particles can be trapped by a combination of static electric and magnetic fields (Penning trap) or by a combination of a static and a radiofrequency (rf) electric field (Paul trap). For a review on ion traps, see e.g. [39]. An advantage of rf traps compared to Penning traps is the absence of magnetic fields, which leads to Zeeman splitting and broadening, and the low attainable ion velocities.

We consider here linear rf traps [40], typically consisting of four cylindrical electrodes, each sectioned longitudinally into three parts (Fig. 1.2, upper part). Compared to the rf quadrupole trap, the linear rf trap offers a line of vanishing rf field, and thus the possibility of storing a larger number of particles with little micromotion. In addition, it also allows good optical access, which is favorable for laser cooling and for spectroscopic measurements.

Radial confinement of charged particles is achieved by applying a radio frequency voltage $\Phi_0 = V_0 - V_{RF}\cos(\Omega t)$ to two diagonally opposite electrodes, with the other two electrodes at ground. $V_{RF}$ and $\Omega$ are the amplitude and the frequency of the rf driving field. In the following we consider mainly the case $V_0 = 0$ (a nonzero value is considered in Section 1.5.1). Axial confinement along the symmetry ($z$) axis of the trap is achieved by electrostatic volt-ages $V_{EC}$ applied to the eight end sections (endcaps). In Fig. 1.2 (lower part) an overview of one complete setup is given [41].

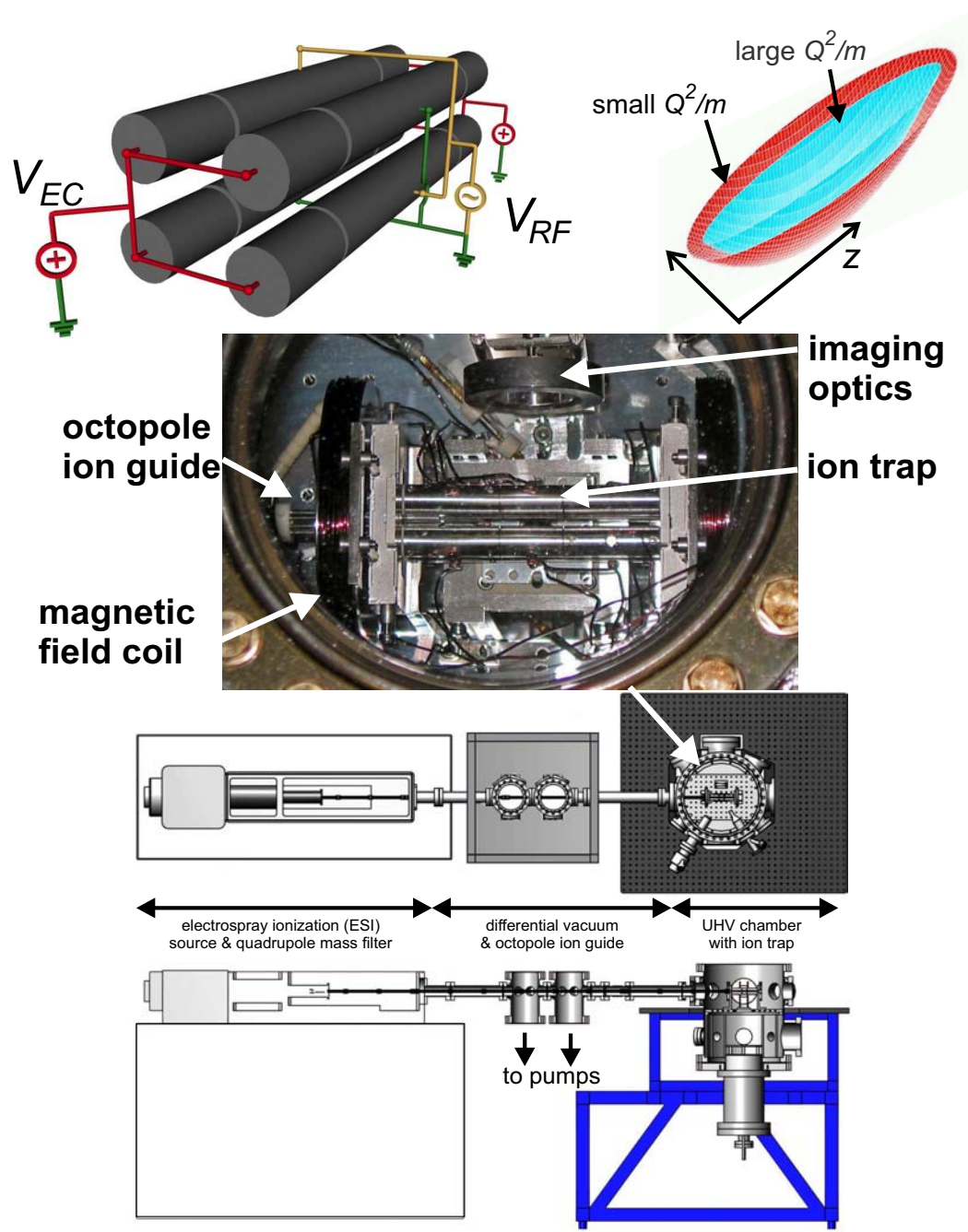

Figure 1.2: Upper part: The linear radiofrequency trap. Schematic of a linear radio frequency trap (left), effective trap potential for different values of $Q^2/m$ (right), and a photograph of the UHV chamber of the ESI/Ba$^+$ apparatus [41, 42]. $\varrho = \sqrt{x^2+y^2}$ and $z$ are the radial and axial trap coordinates, respectively. Lower part: Apparatus for sympathetic cooling of heavy molecular ions [41].

The equations of motion of a single ion inside a linear rf trap are differential equations of the Mathieu type, which lead to stable or unstable solutions with respect to motion in the $x$-$y$-plane, depending on the trap parameters. A necessary condition for stable trapping of an ion or an ensemble of noninteracting ions is a (Mathieu) stability parameter, $q = 2QV_{RF}/m\Omega^2 r_0^2$, $< 0.9$. Here, $Q$ and $m$ are the ion charge and mass, and $r_0$ is the distance from the trap center to the electrodes.

For a single trapped ion, its trajectory is a superposition of an oscillating motion (micromotion) with frequency $\Omega$ and a slow (secular) motion that can enclose a large area in the trap. The amplitude of the micromotion is proportional to $q$ if $q \ll 1$. From simulations it is known that when many ions are to be stored in the trap, a small $q$ parameter is favorable because heating effects induced by the rf field are less pronounced.

For small $q \ll 1$ one can approximate the interaction of the ions with the trap by neglecting micromotion altogether and introducing an effective, time-independent, harmonic potential (quasipotential, pseudopotential)

$$U_{trap}(x,y,z) = \frac{m}{2}(\omega_r^2(x^2+y^2) + \omega_z^2 z^2), \tag{1.1}$$

where $x$, $y$ are coordinates transverse to the trap axis and the electrode center lines cross the $x$ and $y$ axes. The mass-specific motion of a single particle in such a potential is a superposition of harmonic motions with a transverse (to the $z$-axis) oscillation frequency $\omega_r = (\omega_0^2 - \omega_z^2/2)^{1/2}$, with $\omega_0 = QV_{RF}/\sqrt{2}m\Omega r_0^2$ and a longitudinal frequency $\omega_z = (2\kappa Q V_{EC}/m)^{1/2}$. Here $\kappa$ is a trap-geometry constant. An ensemble of (interacting) ions in such a potential has, at sufficiently low (but not too low) average ion energy (i.e. temperature), an approximately constant density and a spheroidal shape [43]. $r_0 = (4.32, 4.36)$ mm, $\kappa \approx (1.5, 3) \times 10^{-3}$ mm$^{-2}$, and $\Omega = 2\pi \cdot (14.2, 2.5)$ MHz for our Be$^+$ and Ba$^+$ trap, respectively.

### 1.3.2 Ion cooling

A variety of atomic ions can be efficiently laser-cooled to mK temperatures and are in principle suitable as coolants.

In our work, beryllium and barium ions are employed. They are produced by evaporating neutral atoms from an oven and ionizing them in-situ in the trap by an electron gun. For laser cooling of Be$^+$ ions light resonant with the $^2S_{1/2}(F=2) \to ^2P_{3/2}$ transition at 313 nm is used [44]. Population losses by spontaneous emission to the metastable ground state $^2S_{1/2}(F=1)$ are prevented by using repumping light red-detuned by 1.250 GHz. $^{138}$Ba$^+$ ions are laser cooled on the $6^2S_{1/2} \to 6^2P_{1/2}$ transition at 493.4 nm [45]. A repumper laser at 649.8 nm prevents optical pumping to the metastable $5^2D_{3/2}$ state. The laser-induced atomic fluorescence can be recorded using photomultiplier tubes and charge-coupled device (CCD) cameras. The latter takes pictures with typical exposure times of $0.5 - 2$ s. The direction of observation is perpendicular to the trap symmetry axis. Because the ions do not "cast a shadow", the CCD images are projections of the complete ion ensemble onto a plane parallel to the trap axis.

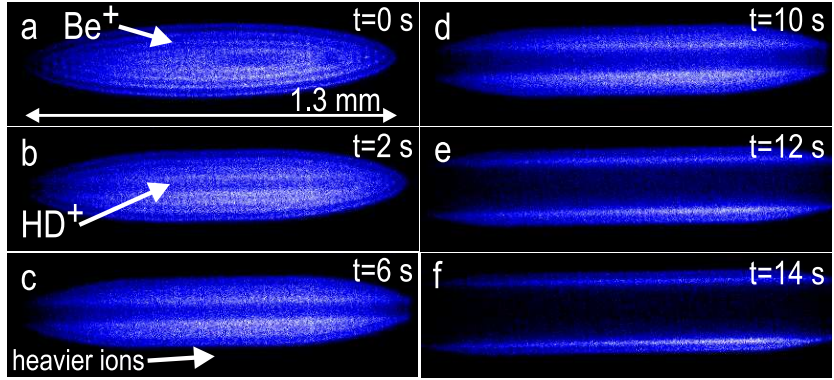

Figure 1.3: Loading sequence of HD$^+$ ions into a cold ($\approx$20 mK) Be$^+$ ion crystal. The presence of cold HD$^+$ ions is obvious from the appearance of a dark (non-fluorescing) crystal core in the initially pure crystal. Heavier atomic and molecular ions were also loaded to the crystal. Since they are less tightly bound by the trap, they are located outside the fluorescing Be$^+$ ensemble and cause its flattening.

Strong cooling results in a transition from a gaseous to a fluid and then to a crystalline state (Coulomb crystal); see Fig. 1.6. The precise meaning of "crystalline" will be explained in Sec. 1.4. The shape of these ensembles is spheroidal if the quasipotential has axial symmetry, Eq. 1.1.

### 1.3.3 Molecular ion production

There are different ways to produce atoms and molecules to be sympathetically cooled. One way is to leak neutral gases into the vacuum chamber and ionize them *in situ* by an electron beam crossing the trap center [46, 47]. The loading rate is controlled by the partial pressure of the neutral gas and the electron beam intensity.

Fig. 1.3 illustrates the loading of HD$^+$ ions into a Be$^+$ ion ensemble prepared beforehand. The shape of such a mixed-species ensemble can be changed in real time by variation of the ratio of radial to axial frequency via variation of the trap parameters, as shown in Fig. 1.4.

The transfer of charged large molecules into the gas phase is possible by several methods. One such method is electrospray ionization (ESI), shown in Fig. 1.5. This well-developed method is capable of producing molecular (multiply) charged beams starting from molecules in solution, even for very massive molecules (several 10,000 amu). Charged droplets of the solution are ejected by a needle and undergo rapid evaporation into smaller and smaller droplets until a final Coulomb explosion causes a significant fraction of single molecules to be created with a variable number of protons attached, originating from the solvent. Molecules of a desired charge-to-mass ratio are then selected using a quadrupole mass filter. Fig. 1.5 illustrates the principle of electrospray ionization and shows a mass-to-charge spectrum obtained for complex molecules. Since the molecules are produced outside the ion trap, a radio frequency octopole ion guide is used for their transport and injection into the trap, see Fig. 1.2 (lower part) [41].

Many other molecular ion species cannot be produced by loading and ionization of neutral gases or by using an ESI source. Here chemical reactions can be used for their production [48, 49, 45], and are described in Sec. 1.6.

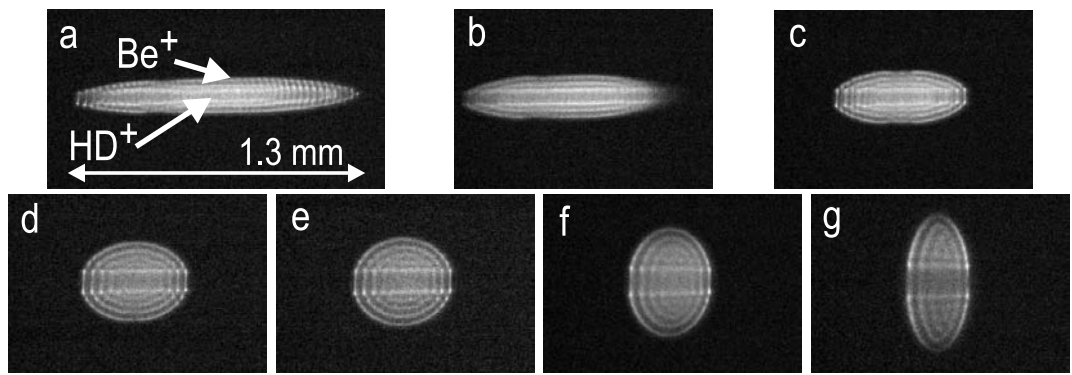

Figure 1.4: Variation of a Be$^+$-HD$^+$ Coulomb crystal shape with trap anisotropy. The static voltage V$_{EC}$ on the endcap electrodes (determining the axial confinement) was changed in (a-c), whereas the amplitude of the rf drive V$_{rf}$ (determining the radial confinement) was changed in (d-g). The pictures are part of a sequence recorded over $\approx$1 minute. The symmetry of the trap potential is cylindrical in all cases. The slight asymmetry in axial direction is due to the laser pressure.

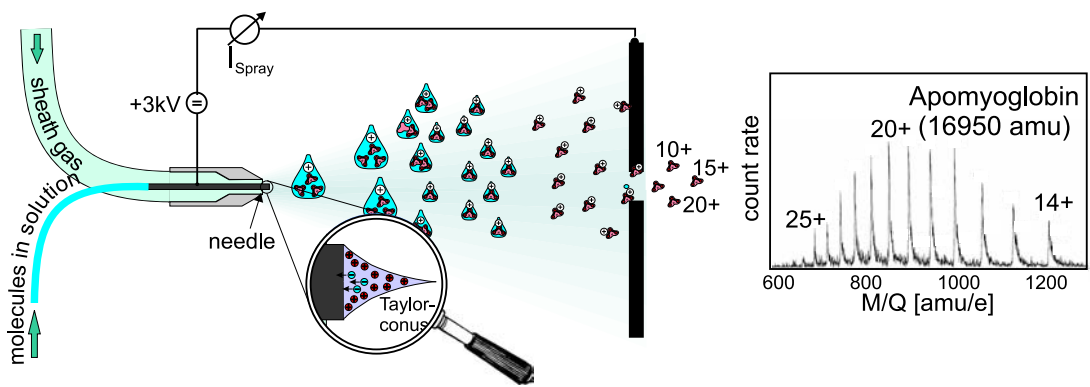

Figure 1.5: Principle of electrospray ionization and a mass-to-charge ratio spectrum of a protein [41]. Values given are protonation charge states.

## 1.4 Properties of cold trapped Coulomb clusters

### 1.4.1 Molecular dynamics simulations

Although an ion ensemble is a classical system (except in the special case of extremely low temperatures, which shall not be considered here), the large number of degrees of freedom, the finite size and the nonlinear ion-ion interaction make analytical treatments very difficult. Fortunately, molecular dynamics (MD) simulations provide an excellent tool for their analysis. They can even be used to analyze novel methods before implementing them ex-perimentally. MD simulations of three-dimensional cold trapped charged particle ensembles have been performed for the one-component plasma in a conservative harmonic potential since the mid-1970s and in rf traps since the first experimental studies [50, 51].

In MD simulations, Newton's equations of motion are solved for all ions in the trap: $m_i \ddot{r}_i = F_i$, where $i$ runs over all ions; $m_i$ and $r_i$ are the mass and position of the ion $i$. The total force $F_i$ acting on each ion arises from individual contributions, some of which depend on ion positions, ion velocities and explicitly on the time $t$,

$$F_i = F_i^{trap} + F_i^{Coulomb} + F_i^{stochastic} + F_i^{laser} \qquad (1.2)$$

with the trap force $F_i^{trap}$, the Coulomb interaction force due to all other ions $F_i^{Coulomb}$, the stochastic force $F_i^{stochastic}$ (due to interactions of ions with the environment, such as collisions with residual gas, scattered light, electric field noise), and the laser cooling force $F_i^{laser}$, which acts on the laser-cooled (LC) ions only and also includes the light pressure force. In the simulations, individual photon absorption and emission processes need usually not be taken into account, since the recoil energy of, for example, $Be^+$ is $k_B(11\ \mu K)$, i.e. 1000 times less than typical temperatures [47, 52, 53].

The MD simulations are a flexible and efficient tool: the effect of various forces can be studied individually or in conjunction with other forces. In particular, the simulations can be performed with the time-varying rf potential, or in the pseudopotential approximation and the results compared. For many purposes the simulations need to compute the ion dynamics only for a few tens of ms. Therefore, for not too large ion numbers ($\sim$1000) the time required to obtain useful information using a personal computer is quite reasonable (hours), making the simulations a practical tool.

MD simulations can be straightforwardly extended to sympathetically cooled ensembles with arbitrary number of species. This was first done in order to explain the experimental structures of a $Mg^+/Ca^+$ mixed crystal [54].

MD simulations are performed in such a way that the ensemble possesses a finite kinetic (secular) energy, equivalent to finite temperature. A basic result is the liquid-crystal phase transition, i.e. the appearance of structure with decreasing temperature, as in Fig. 1.6. In one-species systems, the regimes in which gaseous, fluid and crystallized states occur can be distinguished by the interaction parameter $\Gamma = Q^2/4\pi\epsilon_0 a k_B T$, the ratio between average nearest-neighbor Coulomb energy and thermal energy ($a$: average particle spacing) [43]. With decreasing temperature, an ensemble turns from gaseous to liquid when $\Gamma \geq 2$. In the liquid state, the system exhibits short-range spatial correlations, i.e. the pair-correlation function is non-monotonic. Crystallization sets in at $\Gamma \simeq 170$; it is accompanied by an anomaly of the specific heat [2]. The crystallized state is characterized by "caging" of the ions: their thermal motion amplitude is less than the interparticle spacing, they are nearly localized [55]. These values of $\Gamma$ hold for infinite isotropic systems in a conservative potential [56]. In the approximation where the discreteness of the charges is ignored, the ion spacing or density $n = a^{-3}$ can be derived from the Laplace equation as $n = \epsilon_0 Q V_{RF}^2/(m r_0^4 \Omega^2)$. For typical values $a \approx 30\,\mu$m the phase transition to the Coulomb crystal is then predicted at $\approx 3$ mK. In finite systems, however, the phase transition occurs at lower temperatures, corresponding to larger values for $\Gamma$ [2].

Coulomb crystals do not exhibit the same type of long-range order as solid-state crystals, since the trap potential plays a crucial role. Therefore, the name "crystal" is to be understood as a simplification - the description as a cluster would be more appropriate. Also, in typical experimental situations the ions are not "frozen", but diffuse between crystal sites (see below). In early work, the spatial structures of single-species crystals were investigated [50]. The simulations reproduced the shell structures found experimentally and also showed a regular ordering within shells, a property not directly observable experimentally for large crystals, since so far experiments only image the structures with a single projection image.

At temperatures above a few mK, the case usually encountered experimentally, most ions in a 3D-ensemble are not confined to particular sites, but diffuse between sites; see Fig. 1.7 [2]. Because the CCD images show apparently individual spots, the ensembles seem crystallized, but are not. Except for special sites, the individual spots seen on the experimental images are not the positions where a particular single ion is confined, but where the probability to find any ion is high. Strictly speaking it is thus erroneous (although usual, as here) to denote ensembles at such temperatures as crystallized. "Structured liquids" may be a more appropriate description. Nevertheless, in some sites (e.g. the end sites along the trap axis) the ions are well confined, according to the simulations. For these, their spatial distributions and secular velocity distributions can be obtained, as in Fig. 1.8. This information can be of interest e.g. for determination of the preferred laser propagation direction in spectroscopy and for modelling of the lineshape in spectroscopy.

A particular processes of interest is rf heating, the transfer of energy from the trap field to the ion ensemble. In an important early study [51], it was shown that for a crystallized ensemble, even containing many particles, rf heating is very small. This was confirmed in experiments on $Mg^+$ ions in a ring trap [57], where the Coulomb crystals did not melt even if the cooling light was interrupted for seconds.

A MD study of rf heating in single-species ensembles as a function of trap parameters has been performed in [58, 59]. The influence of phase shift errors on the trap electrodes was investigated in [53] and found not to be important. All these studies indicate that linear quadrupole traps are adequate for cooling and sympathetic cooling of reasonably large ensembles containing thousands of ions; higher-order multipole traps are not a necessity. This statement does not imply that micromotion is irrelevant or rf heating always negligible. In fact, the micromotion kinetic energy is typically a few orders of magnitude larger than the secular energy.

Fig. 1.9 shows crystals with various sympathetically cooled (SC) atomic and molecular species. The most important overall feature of sympathetically cooled and crystallized ensembles is a radial separation of the species due to their different pseudopotential strength: $U_{trap}$ scales as $Q^2/m$, as in Fig. 1.2, upper right panel. In addition, the interspecies interaction is $\sim Q_1 Q_2$. Thus, for equal charge of all ions, the total potential energy will usually be minimized if the lighter ions are closer to the axis [30, 54]. A radial gap between the species develops. For arbitrary charge ratio, in the limit of cylindrical symmetry (very prolate ensembles), the ratio of outer radius $r_1$ of the lower mass-to-charge ratio ($m_1/Q_1$) ensemble and inner radius $r_2$ of the higher mass-to-charge ratio ($m_2/Q_2$) ensemble is given by $r_1/r_2 = (Q_2 m_1/Q_1 m_2)^{1/2}$ [60]. Fig. 1.11 shows an example.

The interaction between the atomic coolants (LC) and the SC ions will be strong if the two species are in close contact, $r_{LC}/r_{SC} \simeq 1$. Most favorable is the case (for equal charge states) in which the coolants are slightly heavier than the SC ions ($M_{LC}/Q_{LC} \geq M_{SC}/Q_{SC}$), effectively "caging" them ($r_{LC} \geq r_{SC}$). This is the reason for the choice of a light atomic coolant ($Be^+$ being the most suitable) for cooling molecular hydrogen ions and for the choice of a heavy atomic ion for heavy molecular ions. Since atomic ion masses are limited to about 200 amu, the cooling of very heavy molecular ions is disfavored. However, if those ions carry a large charge, cooling is facilitated (see Sec. 1.8). Examples of crystals with radial separation can be seen in Figs. 1.3, 1.9, 1.10, and 1.11. Particularly interesting is the "tubular" structure that ensembles with three or more species can exhibit; see Fig. 1.9 c.

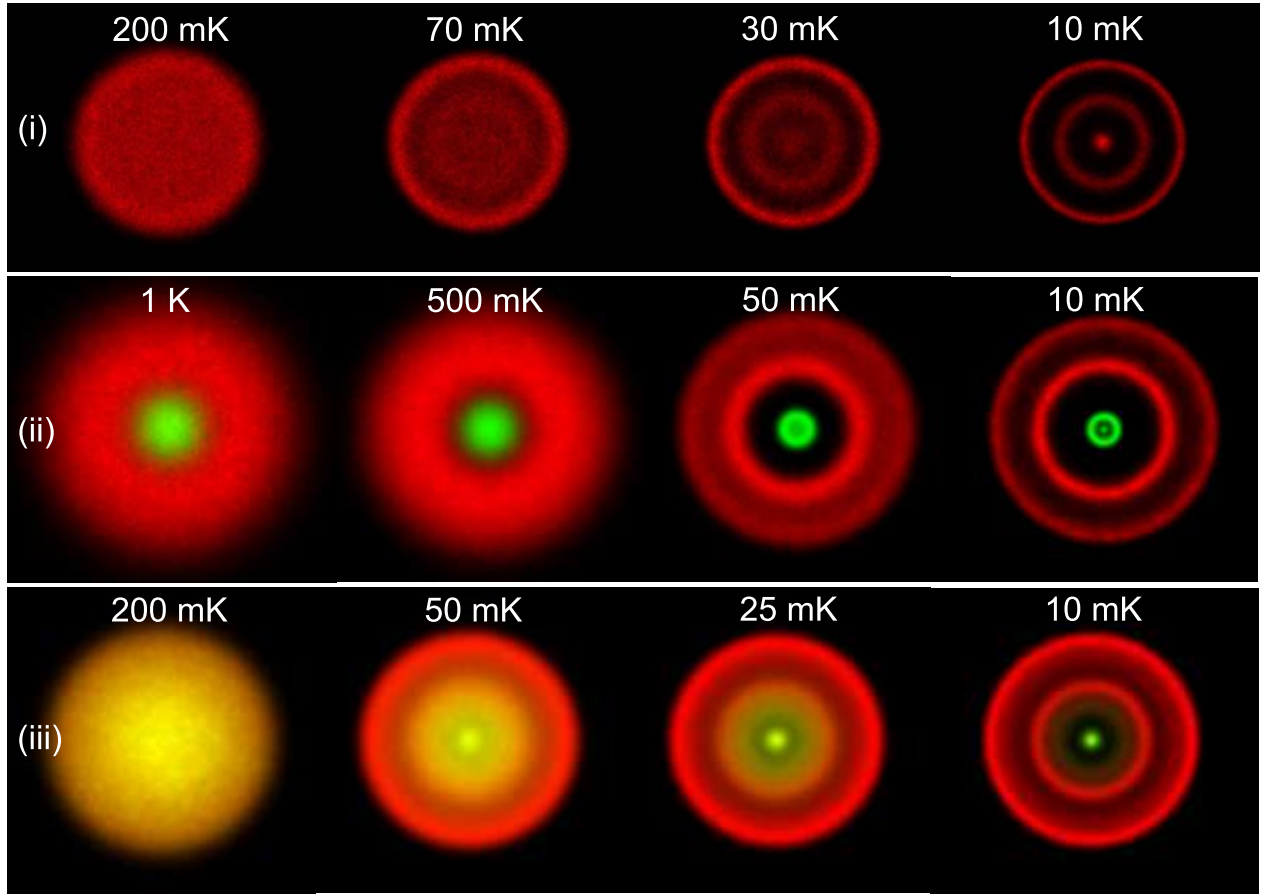

Figure 1.6: Fluid-to-crystal phase transition. MD simulation of ensembles with (i) one species (500 Be$^+$ ions), (ii) two species (500 Be$^+$ ions (outer shells) and 100 HD$^+$ ions (inner shells)), and (iii) two species (500 Ba$^+$ ions (outer shells) and 100 barium isotope ions (inner shells)) at different translational temperatures. Particle trajectories integrated over 1 ms are displayed. The view is in the $x$-$y$-plane. Only a short section along $z$ is shown. Clear separation between the ion shells is observed at temperatures $\leq$30 mK, while the ensembles are in the fluid state above $\approx$200 mK. The shell structure develops around 200 mK. With lowering of the temperature, the radial "sharpness" of the shells increases. The ions are fully localized.

Since the radiation pressure of the cooling laser does not act on the SC ions an asymmetry in axial direction occurs if a single axial cooling laser is employed. The SC ions are the located closer to the laser (Figs. 1.9 b, 1.12 a). This feature is particularly apparent when the number of SC ions is significantly smaller than the number of the atomic coolants.

One important application of the simulations to experimental observations is the determination of ion numbers and translational temperatures. Fig. 1.12 is a simple example.

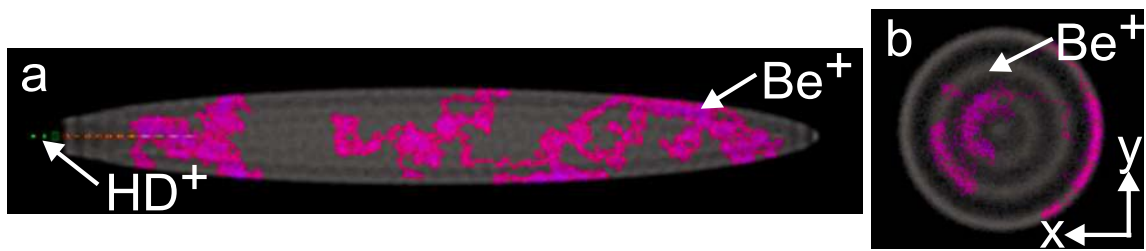

Figure 1.7: Ion diffusion in Coulomb crystals. (a) Trajectories of several individual Be$^+$ ions in a cold (10 mK) Be$^+$-HD$^+$ ion crystal (time-averaged trap potential assumed; duration: 1 ms). Because of diffusion, except for special sites, individual spots in a (simulated) CCD image are not the positions where a particular single ion is confined, but where the probability to find any is high. (b) Axial view of a section of the crystal in (a).

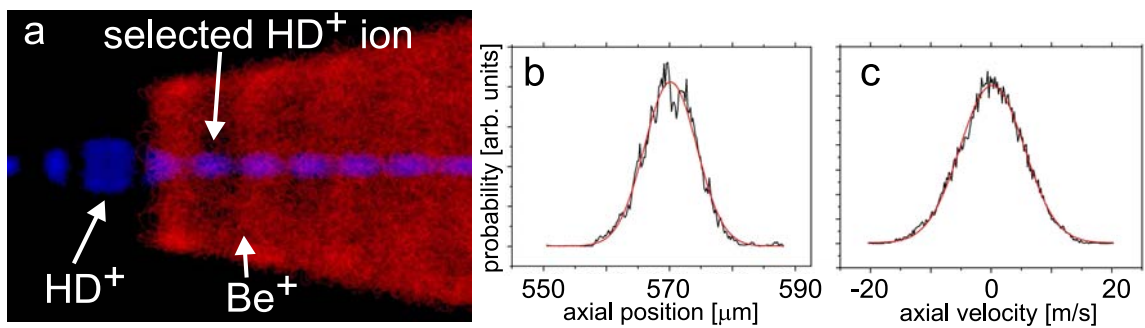

Figure 1.8: An individual, localized HD$^+$ ion at 10 mK (a) and (b,c) its spatial ($z$) and velocity ($v_z$) distribution. The ion is embedded in a (10 mK) Be$^+$ crystal. The particle trajectories were simulated for 1 ms. Smooth lines are Gaussian fits.

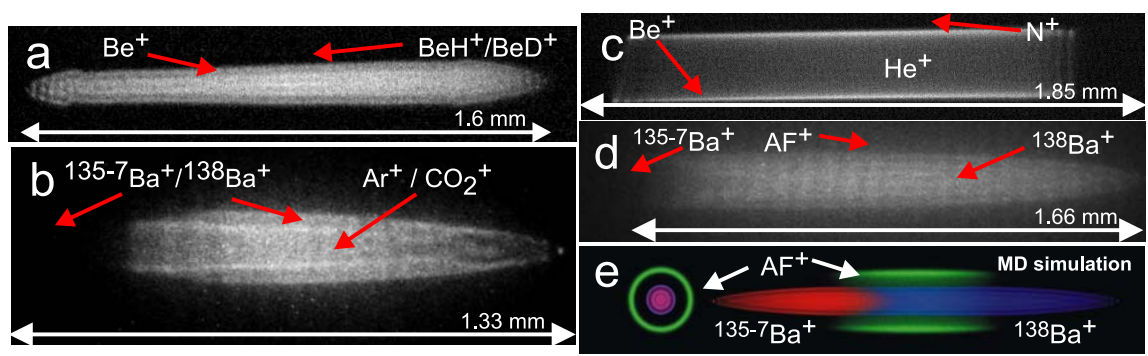

Figure 1.9: Examples of mixed-species crystalline ion plasmas. (a) 900 Be$^+$, 1200 BeH$^+$ and BeD$^+$ (in equal amounts) at 15 mK [48]. (b) 300 $^{138}$Ba$^+$, 150 $^{135-137}$Ba$^+$, 240 BaO$^+$, and ≈200 Ar$^+$ and CO$^+$ at ≈20 mK [45]. (c) 500 Be$^+$, 1500 He$^+$, and 800 N$^+$ at ≈12 mK (d) 830 $^{138}$Ba$^+$, 420 $^{135-137}$Ba$^+$, and 200 protonated Alexafluor ions (AF$^+$, mass 410 amu). The temperature of the Ba$^+$ including SC isotopes is at 25 mK, the AF$^+$ are at ≈88 mK. (e) MD simulation of the crystal in (d) [52].

Simulations are performed for various temperature values and various ion numbers and the best match to the experimental image yields the respective values. For medium-size crystals ($< 500$ ions) containing mostly LC ions, the inaccuracy in the number determination is a few percent only; see Fig. 1.13 [53]. The resolution of the temperature determination via MD simulation was higher than a direct measurement via the atomic lineshape; see Fig. 1.12.

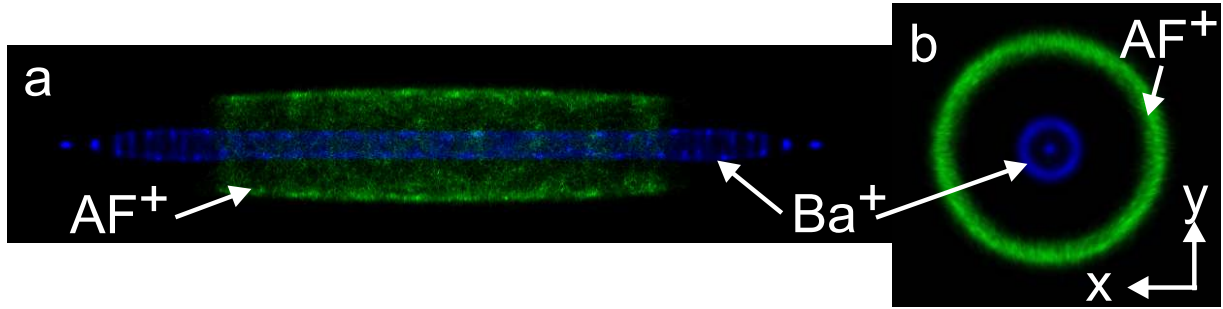

Figure 1.10: Spatial separation of species. (a) Simulation of an ion crystal containing 50 laser-cooled (LC) Ba$^+$ ions and 50 sympathetically cooled (SC) AF$^+$ ions at 15 mK and 18 mK, respectively. Here, rf micromotion is included. (b) Cross section of the crystal in (a). The micromotion is directed towards the electrodes (located on the $x$ and $y$ axes) and leads to a (slight) blurring of the AF$^+$ ions in radial direction, in addition to the blurring caused by secular motion and found also when the simulations are performed in a time-averaged pseudopotential; see [53].

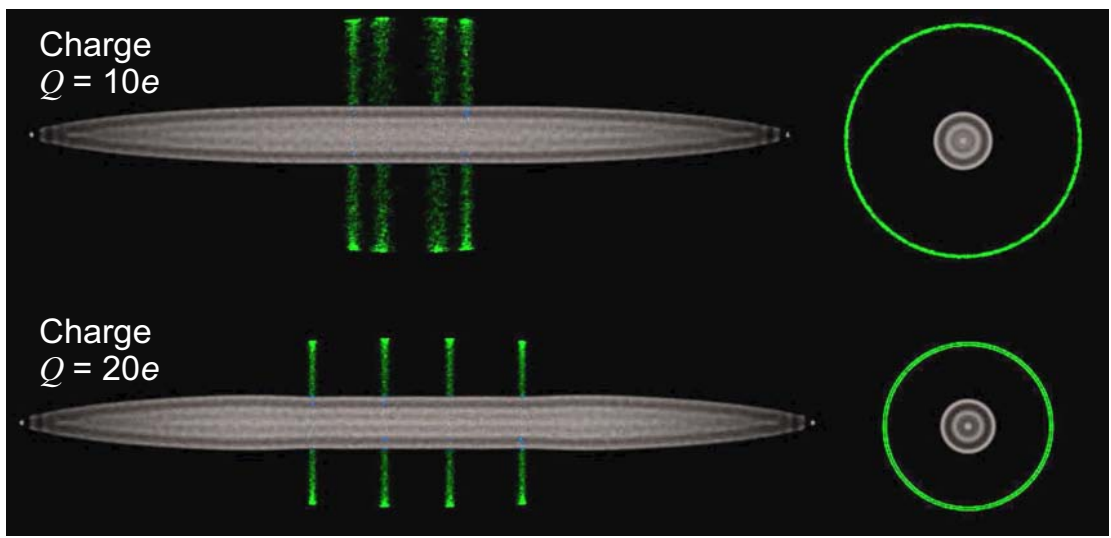

Figure 1.11: MD simulation of a Coulomb crystal containing 1000 Ba$^+$ ions and 20 highly charged molecular ions (16000 amu), of charge $Q_{SC} = 10\,e$ (top) and $20\,e$ (bottom).

### 1.4.2 Collisional heating of ion crystals

An interesting aspect is the stability of ion crystals under (elastic) collisions with a neutral gas. Such collisions can lead to heating which can be substantial. Here, the most important ion-neutral interaction arises from an induced dipole attraction with potential $\varphi = -(\alpha/2)(e/(4\pi\epsilon_0 r^2))^2$ for a singly charged ion [61]. $\alpha$ is the polarizability of the neutral atom or molecule, $e$ the electron charge, and $r$ the radial separation. Using classical collision theory, one can derive expressions for the ion-neutral collision heating (or cooling) rate $h_{coll}$ and the momentum transfer collision rate $\gamma_{elastic}$, respectively [61]:

$$h_{coll} = \frac{3 \cdot 2.21}{4} \frac{ek_B}{\epsilon_0} n_n \sqrt{\alpha\mu} \frac{T_n - T_c}{m_n + m_c}, \qquad \gamma_{elastic} = \frac{2.21}{4} \frac{e}{\epsilon_0} n_n \sqrt{\frac{\alpha}{\mu}}, \qquad (1.3)$$

where $n_n$ is the particle density of the neutral gas, $m_n$ ($m_c$) and $T_n$ ($T_c$) are the masses and temperatures of the neutral (charged) particles, $\mu = m_n m_c/(m_n + m_c)$ is the reduced mass.

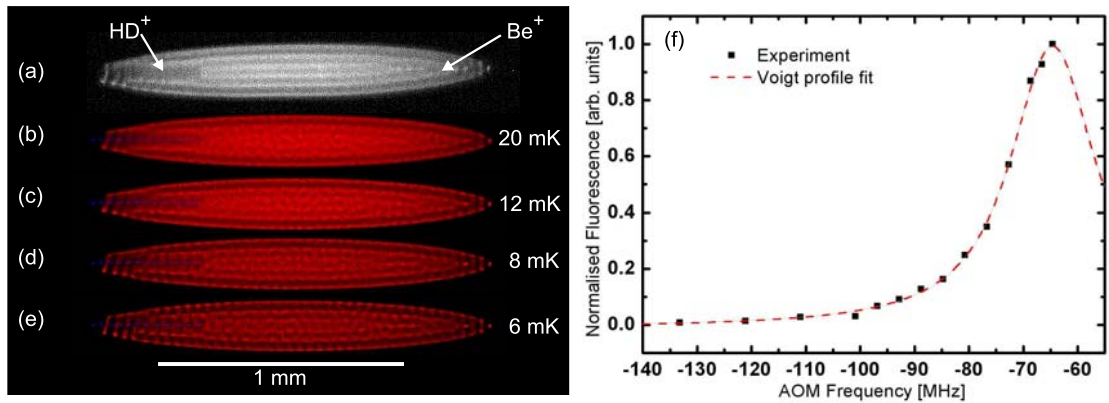

Figure 1.12: Determination of translational temperatures of LC and SC ions [53]. CCD image (a) and simulated images (b-e) of a two-species ion crystal are shown on the left. Ion numbers (690 Be$^+$ and 12 HD$^+$ ions) and translational temperatures (10 mK) for LC and SC ions are obtained by visual comparison between experimental and MD images [47]. The asymmetry of shape in axial direction is caused by light pressure forces. (f) Determination of the translational temperature of a pure Be$^+$ ion crystal by fitting a Voigt profile to the fluorescence line shape yielding a temperature of $5 \pm 5$ mK. The fluorescence line shape was obtained by scanning the cooling laser frequency over the Be$^+$ resonance using an acousto-optical modulator (AOM). Note, that the AOM frequency values have an arbitrary offset compared to the atomic resonance.

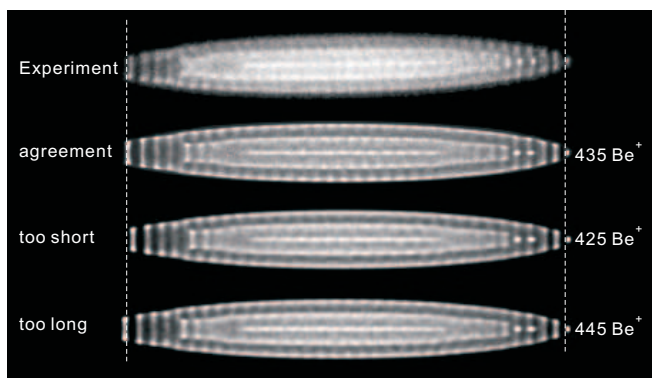

Figure 1.13: Determination of ion numbers. The experimental image of a Be$^+$ ion crystal is compared with simulated images of ensembles with different ion numbers. The best fit is achieved for a number of 435 ions [53].

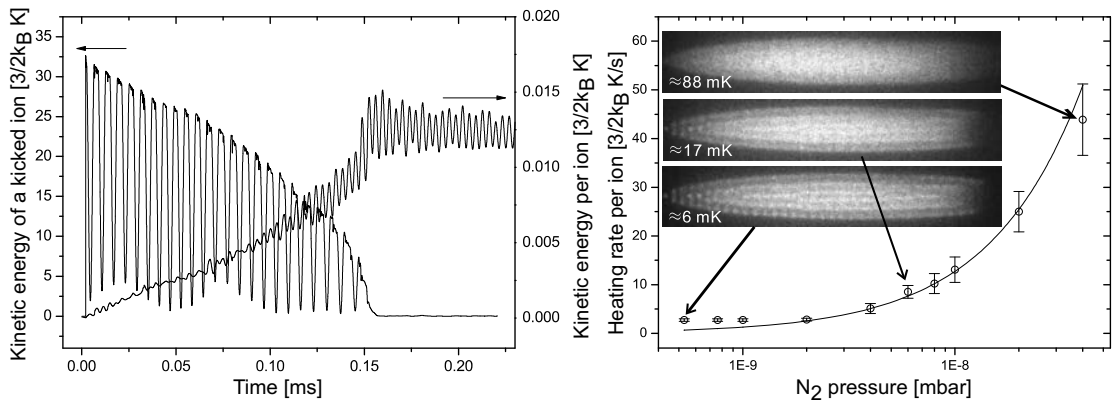

Figure 1.14: Left: Ion-neutral head-on collision. After its collision with a neutral helium atom a barium ion gradually transfers its gained kinetic energy (left scale) to the whole ion ensemble, whereupon the kinetic energy per ion (right scale) increases (simulation was performed without micromotion). Right: Collision heating of a $Ba^+$ ion crystal as a function of nitrogen pressure. Line: Eq. (1.3), symbols: heating rates deduced from the experiment using MD simulations [53].

For example, in a background gas of $N_2$ at 300 K and a pressure of $1 \cdot 10^{-9}$ mbar the average $Ba^+$-$N_2$ elastic collision rate is $\gamma_{elastic} \approx 0.017$ s$^{-1}$ per ion. In each collision the average energy transfer is $\approx k_B$ (128 K), leading to a heating rate $h_{coll} = k_B(2.2$ K/s$)$ per ion, which increases linearly with the residual gas pressure. This should be compared with the heating rates found for our traps and ensembles, Tab. 1.2.

The consequences of the collision between an ion and a residual gas atom or molecule can be simulated. Fig. 1.14(left) shows the kinetic energies of an ion ensemble of 1249 barium ions and of a single ion after the latter has suffered a head-on collision with a helium atom of $\frac{3}{2}k_B(300$ K$)$ kinetic energy. The colliding barium ion suddenly gains a large velocity (76.8 m/s) leaving the ion crystal. It starts to oscillate in the trap. Its kinetic energy is periodically transformed into potential energy and back. Each time the ion passes through the ion cluster, it transfers some energy to it, with the energy loss rate being lower, the faster the ion is [62]. Finally, the initial kinetic energy is distributed to all ions and the whole ensemble reaches an equilibrium state with increased potential and kinetic energies per ion. For a typical crystal considered here, the thermalization time is $\tau \sim 0.2$ ms. Simulations including micromotion indicate that also some micromotion energy is converted into secular energy, increasing the heating rate above the value given by Eq. 1.3.

The collision heating of an ensemble of $N$ ions will be essentially continuous if the time between two collisions in the whole ensemble, $1/N\gamma_{elastic}$, is smaller than the thermalization time $\tau$, i.e. for ion numbers $N > 1/(\gamma_{elastic}\tau)$. For a typical value of $\gamma_{elastic} = 0.002$ s$^{-1}$ per ion at $1 \cdot 10^{-10}$ mbar, this condition is not fulfilled for typical size ensembles, but for ensembles with $N > 2500000$ ions. In fact, if the ensemble is small and observed for a time much shorter than the mean interval between collisions, $1/(N\gamma_{elastic})$, the ensemble may appear colder than the time-averaged temperature. With a typical CCD exposure time of 2 s this is the case for $N < 250$ ions. Indeed, our experience is that small clusters show a lower temperature (as found from the CCD images) than larger ones.

For large ion ensembles ($N > 2000$) and sufficiently long CCD camera exposure times there is a sufficiently large total number of collisions in the ensemble, that their effect can be modelled in a very simple way. Frequent velocity kicks per ion, but with very low velocity kick magnitudes are implemented. In this way computing time can be saved because equilibrium states are reached much faster (within a few ms of simulated time).

The heating of an ensemble of Ba$^+$ ions at different N$_2$ pressures was studied experimentally; see Fig. 1.14. Assuming a realistic cooling rate $\beta = 2 \cdot 10^{-22}$ kg/s [53], at each pressure the heating rate $h_{coll}$ that yields the best agreement of simulated and experimental images was fitted. Here, $T_n - T_c$ (Eq. 1.3) was set to 300 K. For pressures above $1 \cdot 10^{-9}$ mbar, the experimental values agree well with the theoretical heating rate Eq. 1.3, with $\alpha_{N_2} = 4\pi\epsilon_0(1.76 \cdot 10^{-24}\,\text{cm}^3)$.

At the lowest pressures the heating appears to reach a pressure-independent level. This may in part be an artefact due to the finite spatial resolution of our imaging system (which leads to blurring of the crystal image). There may also be heating sources not taken into account, such as electric field noise. As they are hard to quantify, they are not implemented in our simulations directly, but their effect can be included ad hoc in the velocity kick model as a pressure-independent contribution to the kick magnitude [53].

### 1.4.3 Heating effects

The sympathetic cooling is exerted by the LC particles via the Coulomb interaction and depends on the spatial distributions, the temperatures and the ion numbers of the LC and SC species. Heating of the LC particles competes with sympathetic cooling. A heating rate $h_j = dE_j/dt$ characterizes the rate of increase of total energy $N_j E_j$ (potential, secular kinetic and micromotion kinetic) of a sub-ensemble $j$ due to the combined effects of trap noise, collisions with background gas and rf heating. It depends on the spatial distribution of the ion sub-ensemble and is expected to increase with increasing distance from the trap axis, since the micromotion amplitude increases linearly with radial distance. Assigning a single heating rate to a subensemble is an approximation, since the subensemble typically has a finite radial extent; however, the energy exchange within a subensemble is fast and if slow processes are to be described, we may consider the average heating rate only. Note that it would not be accurate (except in special situations) to assume $h_j = dE_{j,secular}/dt$: as a species is heated, its kinetic energy increases, but so does its spatial extension and therefore its potential energy [53]. Therefore not all heating goes into temperature increase.

For a given LC ion temperature, the SC ion temperature adjusts such that for this sub-ensemble heating and cooling rates are equal in magnitude. In the simulations, it is convenient not to use the LC ion temperature as an input parameter, but instead a LC heating rate $h_{LC}$ and the laser cooling coefficient $\beta$. The laser cooling is modelled by a viscous laser cooling friction force, $\boldsymbol{F}_i^{laser} = (-\beta \dot{z}_i + const)\mathbf{e}_z$, so that the laser cooling rate is itself temperature dependent. In the approximation where the micromotion is negligible, the cooling rate $c_{LC}$ is proportional to the secular kinetic energy and thus the temperature:

$$c_{LC} = \left(\frac{dE_{LC}}{dt}\right)_{lasercooling} \simeq -\frac{\beta}{m_{LC}} k_B T_{LC}. \tag{1.4}$$

Laser cooling $\beta$, coolant heating $h_{LC}$ and the sympathetic heating exerted by the SC ions on the LC ions determine the LC ion temperature.

Consider first the ensemble of Fig. 1.12, where a simplified situation occurs. Because the number of SC ions is very small and they are embedded, as a first approximation, a common SC and LC heating rate is set, and a realistic laser cooling coefficient. Three parameters, heating rate and two ion numbers, are varied until the simulated image of the LC ion ensemble agrees with observation. This gives an LC temperature of 10 mK. Then the SC heating rate is varied independently, allowing to study its effect onto the simulated structure. When its value is sufficiently high, the sympathetic heating causes the shell structure of the LC ensemble to disappear. This occurs when the SC temperature is at 50 mK. By requiring that the structure of the neighborhood of the SC ions agrees with observation, an upper limit for the SC heating rate is found, with a corresponding upper limit of the SC ion temperature, $T_{SC,max} \simeq 20$ mK. The accuracy of the SC temperature determination is of course limited by the quality of the experimental image due to the finite spatial resolution of the imaging optics. It is likely that the SC temperature is nearly equal to the LC temperature.

An extreme situation in sympathetic cooling occurs when the SC particles are significantly heavier than the LC particles (and of equal or larger charge). Then the large spatial separation reduces the interaction strength and thus the cooling power. A significant difference in the temperatures of the two species can arise. This is unlike the above case where the SC particles are lighter than the LC ions and thus well-embedded, leading to an efficient coupling. From an experimental point of view this situation can be difficult to characterize in a straightforward way via the CCD images, since for a large spatial separation and a small SC ion number, their effect on the shape of the LC ensemble is not easily observable. For such a situation the simulations are an important tool. As one example, the sympathetic cooling of protonated Alexafluor ions (AF$^+$, mass 410 amu) by $^{138}$Ba$^+$ (Fig. 1.9 d) could be characterized with the help of the simulation, Fig. 1.9 e. The ensemble actually contains five species, of which one species is laser-cooled and the other four are sympathetically cooled. Three of the SC species are Ba$^+$ isotopes different from $^{138}$Ba$^+$; they are grouped together as a single SC species (SC2). In order to model a three-species system, we must introduce three heating rates $h_{LC}$, $h_{SC1}$, $h_{SC2}$. In steady state, energy conservation requires that the heating rates and $^{138}$Ba$^+$ temperature $T_{LC}$ are related by:

$$dE_{tot}/dt = (-\beta k_B T_{LC}/m_{LC} + h_{LC})N_{LC} + h_{SC1}N_{SC1} + h_{SC2}N_{SC2} = 0. \quad (1.5)$$

To obtain insight into the effect of the Coulomb interactions between the species, Fig. 1.15 shows a MD simulation in which the various heating and cooling interactions are "switched on" in time to clarify their effect. The middle section of the plot indicates that in absence of heating of the two SC species their temperature is equal to that of the LC particles. The unavoidable heating of the two SC species, turned on in the right section of the plot, increases the temperatures and leads to different temperatures for the three species, due to the differing interaction strengths caused by the different spatial distributions. The simulation also shows that the response time of the sub-ensembles is on the order of 10 ms.

Returning to the characterization of the experimentally obtained ensemble, the main goal was the determination of the temperature of the heavy SC species. The experimental data consisted of the Ba$^+$ cooling coefficient $\beta/m_{Ba} \approx 760$/s, the CCD image of the (effective) three-species ensemble, the number of heavy SC1 ions, and the CCD image of the (effective) two-species ensemble, where the latter two data were obtained by emptying the trap of the heavy SC ions. The two CCD images were rather similar, which indicated a relatively weak LC-SC interaction compared to the LC heating effects. With these inputs, the MD simulations were employed to determine four quantities: the number of LC and SC2 ions, the LC heating rate (with the reasonable assumption that the heating rate for the SC2 Ba$^+$ isotopes has the same value), and the SC1 heating rate for the heavy ions. In fact, the two-species simulation fit of the CCD images of the system without AF$^+$ easily yields a temperature $T_{LC,0}$ and the ion numbers $N_{LC,0}$, $N_{SC2,0}$, from which by Eq.(1.5), the common heating rate $h_{LC} = (\beta/m_{LC}) k_B T_{LC,0} N_{LC,0}/(N_{LC,0} + N_{SC2,0})$ is obtained. The ion numbers $N_{LC,0}, N_{SC2,0}$ could be obtained rather directly, since their determination is not very sensitive to the SC1 heating rate. Finally, the three-species simulation was then performed assuming the heating value $h_{LC}$ remains unchanged in the presence of AF$^+$. This assumption is reasonable, since the change in size of the LC/SC2 ensemble was small. Thus, only the heating rate $h_{SC1}$ was varied, until the CCD image was reproduced. It is in this step that the interspecies interactions are crucial: the heating of the relatively distant heavy ions affects the laser-cooled ions, heating them to a temperature $T_{LC}$ given by $\beta k_B T_{LC}/m_{LC} = h_{LC}(1 + N_{SC2}/N_{LC}) + h_{SC1} N_{SC1}/N_{LC}$. The temperature value $T_{SC1}$ then follows from the simulation result. In order to estimate its uncertainty, the uncertainty in the CCD image match and in the value of $\beta$ (heating rates and temperatures scale with $\beta$) were considered, leading to a factor of 3 between upper and lower limits for $T_{SC1}$ (i.e. $T_{AF^+}$), the former being 138 mK. The interesting sympathetic cooling rate of the heavy ions SC1 by all other ions (here LC and SC2) is just equal in magnitude to the heating rate $h_{SC1}$, because of equilibrium. Table 1.2 summarizes the various rates for this particular experiment (for a value $\beta/m_{LC} = 760$/s). Energy conservation, Eq.(1.5), is not exactly fulfilled because the value $T_{LC}$ is obtained from the simulations with a certain inaccuracy.

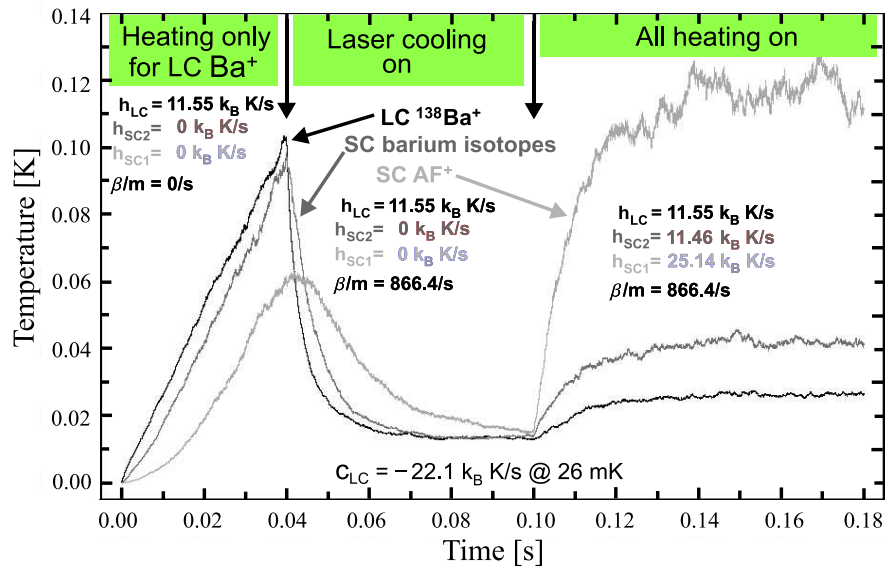

Figure 1.15: MD simulation of sympathetic heating and cooling in a multi-species ion ensemble. At $t = 0$ all species were set to a secular temperature of 0 K. Left: heating is initially implemented only for the $^{138}$Ba$^+$ ions (heating rate $h_{LC} = 11.55 k_B$ K/s). The other ions are sympathetically heated. Center: at $t = 0.04$ s laser cooling of $^{138}$Ba$^+$ is switched on (cooling coefficient $\beta/m_{LC} = 866.4$/s), and the $^{138}$Ba$^+$ temperature decreases. The $^{138}$Ba$^+$ ions now sympathetically cool the other ions, and the ensemble reaches an equilibrium state at nonzero temperature. Right: at $t = 0.1$ s, heating is also turned on for the SC barium isotopes (heating rate $h_{SC2} = 11.46 k_B$ K/s) and the SC $^{410}$AF$^+$ ions (heating rate $h_{SC1} = 25.14 k_B$ K/s), which increases the temperatures of all species until they reach the equilibrium state.

Thus, the simulation shows that approx. 20% of the available total laser cooling "power" $c \cdot N_{LC}$ acts on the AF$^+$ ensemble via the Coulomb interaction, and approx. 46% acts on all SC ions together.

| species $j$ | number $N_j$ | temperature $T_j$ [mK] | laser cooling rate $c_j$ [$k_B$ K/s] | heating rate $h_j$ [$k_B$ K/s] | sympathetic rate ($c_j$-$h_j$) [$k_B$ K/s] |
|---|---|---|---|---|---|
| $^{138}$Ba$^+$ (LC) | 830 | 25 | -19 | 9.9 | 10.4 (due to SC1, SC2) |
| $^{410}$AF$^+$ (SC1) | 200 | 88 | 0 | 15.9 | -15.9 (due to LC, SC2) |
| $^{135-137}$Ba$^+$ (SC2) | 420 | 37 | 0 | 9.9 | -9.9 (due to LC, SC1) |

Table 1.2: Summary of some properties of the multi-species system, Fig. 1.9 e. Rates are normalized to the species' particle number.

The comparison between observed and simulated ion crystals allows or requires also the determination of the trap parameters, pseudopotential frequencies and (spurious or deliberately applied) offset potentials. An example is shown in Fig. 1.16. This mixed-species ensemble is not axially symmetric because of static potentials on the electrodes. It also contains several species. Even in such complicated cases, MD simulations can provide a good explanation, after the ion numbers and trap parameters have been fitted by comparison with the CCD image - which shows only one species.

## 1.5 Characterization and manipulation of multi-species ensembles

### 1.5.1 Crystal shapes

The overall shape of a cold ion plasma depends strongly on the symmetry and "shape" of the trapping potential. For axial trap potential symmetry, the ensembles are spheroidal, and have been characterized both in linear rf and Penning traps [32, 63]. In absence of axial symmetry the shape of the plasma has been predicted to be that of an ellipsoid [64]. Ellipsoidal plasmas were first observed and studied in Penning traps [65]. The first demonstration in a linear rf trap was given with our Be$^+$ apparatus, by adding a static quadrupole potential to the trap electrodes, resulting in an fully anisotropic quasipotential [66]. Such a situation can be produced with crystals containing two or more different ion species; see Fig. 1.17. A comparison of experimental shapes to theoretical predictions from a simple cold fluid plasma model shows agreement for small anisotropy, while deviations observed for larger anisotropy can be explained by the presence of the SC particles causing space charge effects.

The ability to reversibly deform cold multi-component crystals by static quadrupole potentials is interesting for several reasons. It allows for (i) a controlled ejection of heavier ion species from the trap (see below), (ii) for a complete radial separation of lower-mass SC ions from the LC ions, and (iii) opens up the possibility of studying trap modes of oscillation of ellipsoidal crystals, in particular of multi-species crystals. Conversely, a precise measurement of the trap modes of oscillation of cold ion crystals allows for the identification of even small anisotropies of the effective trap potential, which is important for precision measurement applications and the characterization of systematic effects, such as offset potentials [45].

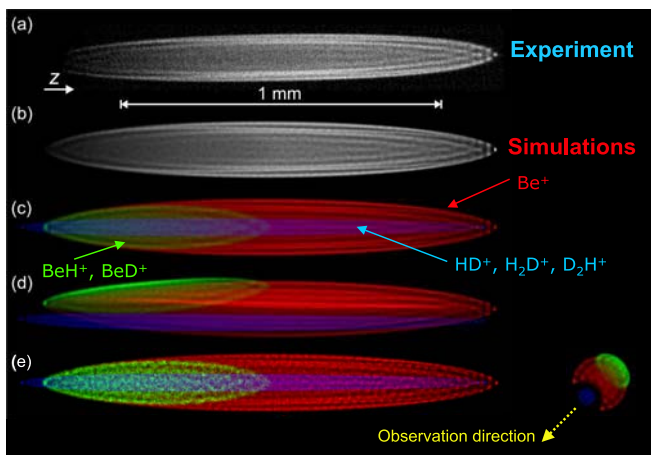

Figure 1.16: MD analysis of an asymmetric cold multi-species ion crystal containing Be$^+$ ions and various SC molecular ion species. (a) CCD image. The view is in the $z$-$y$-plane (as indicated by the arrow in the right panel in (e)). (b-e) MD simulations of the crystal in (a). (b) shows the Be$^+$ only, whereas (c-e) show all species in the crystal. The images show projections of the crystal onto the $z$-$y$-plane, (b,c) and left panel in (e), the $z$-$x$-plane, (d), and the $x$-$y$-plane, right panel in (e), respectively.

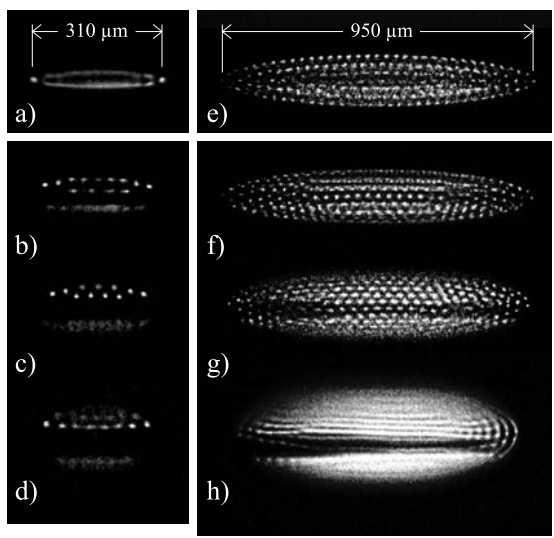

Figure 1.17: Ellipsoidal deformation of mixed-species crystals by a static quadrupole voltage $V_{DC}$ applied to the central electrodes. For $V_{DC} = 0$ the crystals are spheroidal. Left: small crystal containing $\simeq 20$ Be$^+$ and a smaller number of SC impurity ions at different values $V_{DC}$: 0 V (a), 2.8 V (b), 3.6 V (c), 4.2 V (d). Right: medium-sized crystal containing $\simeq 500$ Be$^+$ ions. $V_{DC}$ is set to 0 V (a), 1.4 V (b), 2.8 V (c), 4.0 V (d). Asymmetries in (b)-(d) and (f)-(h) are due to stray electric fields [66].

Furthermore, by applying static offset potentials to individual trap electrodes a spatial manipulation of the ion crystals is possible. For example, by carefully aligning the atomic coolant ions around the dark core containing SC ions (which is possible with an inaccuracy below 20 $\mu$m, while observing the crystal using the CCD camera) trap imperfections can be minimized more efficiently compared to the commonly used fluorescence correlation measurements; see, e.g., [45, 67].

### 1.5.2 Particle identification - destructive and nondestructive

Since fluorescence detection by repeated absorption-emission cycles is not applicable to trapped molecular ions in UHV, i.e. in absence of collisions with a buffer gas [68], different techniques are required for their reliable identification. A commonly used destructive technique for molecular ions is time-of-flight mass spectroscopy. We have used a simplified variant in the Ba$^+$ apparatus. The trapped ions are extracted from the trap by reducing the radio frequency amplitude, in the presence of a finite d.c. quadrupole potential $V_0$, which causes the ion trajectories to become unstable (the Mathieu $q$-parameter enters the instability region). Heavy and hot ions escape first. Upon leaving the trap, the ions are guided to and attracted by the cathode of a channel electron multiplier (CEM) and counted.

Fig. 1.18 shows an example of a mass spectrum obtained from a multi-species ensemble. The ion count peaks occur at clearly separate radio frequency amplitudes, allowing for identification whether LC and SC ions are present. Assuming equal detection efficiencies, the ion signal sizes are used to determine the ratio of the numbers of LC and SC ions. In addition, the spectrum also provides evidence that the heavier molecules are sympathetically cooled by the barium ions, as can be seen by comparing the two insets in Fig. 1.18. Fig. 1.18 b shows the mass spectrum of an ensemble when the Ba$^+$ ions have been laser-cooled. The ion count peaks of both LC and SC ions are narrower, indicating a narrowed energy distribution. An accurate temperature determination is not feasible, however.

A nondestructive detection technique is often desirable. One such technique is based on the excitation of motional resonances of trapped species. The information about the species is encoded in the frequency of the resonance. This technique has been used in the past for mass spectrometry of ion clouds in the gas or fluid state [29, 69]. For a "two-ion crystal" containing a single molecular ion sympathetically cooled by a single atomic ion, a high mass resolution has been demonstrated [31].

Atomic and molecular species identification via motional resonance-based detection is also possible in more complex systems, namely in large multi-species ion crystals of various size, shape and symmetry [47, 52, 70]; see Fig. 1.19. The basic principle of the method is as follows: the radial motion of the ions in the trap is excited using an oscillating electric field of variable frequency applied either to an external plate electrode or to the central trap electrodes. When the excitation field is resonant with the oscillation mode of one species in the crystal, energy is pumped into the motion of that species. Some of this energy is distributed through the crystal, via the Coulomb interaction. This, in turn, leads to an increased temperature of the atomic coolants and modifies their fluorescence intensity, which can be detected.

### 1.5.3 Motional resonance coupling in cold multi-component ion crystals

According to the quasipotential approximation, the radial oscillation frequency of noninteracting trapped ions $\omega_r \simeq \omega_0 \propto Q/m$ (Eq. 1.1) when the axial potential is weak, i.e. the ensembles are strongly prolate. Then, the ratios of motional frequencies of different singly-charged species are equal to the mass ratios. This is indeed what is observed when the ensemble is in the gas state, where the density is low and therefore the ions are weakly interacting. Interactions between the different ion species, especially in the crystalline state, can shift the observed motional frequencies by a significant amount. This complicates the analysis of the experimental spectra, in particular, for mixed-species ion crystals with SC ions having comparable mass-to-charge ratios. Examples of spectra obtained for such an ion crystal are displayed in Fig. 1.20.

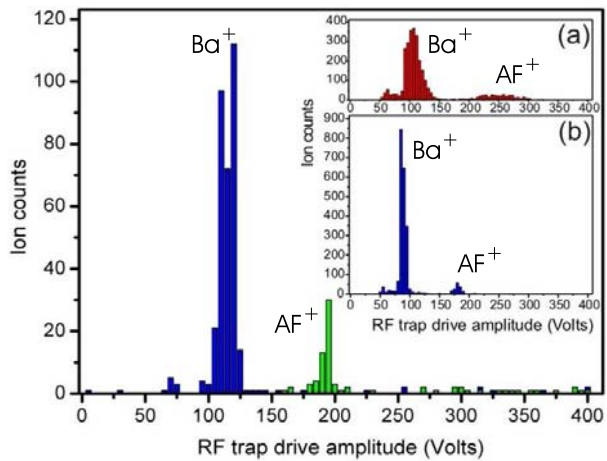

Figure 1.18: Extraction of ions (of crystal in Fig. 1.9(d)) from the trap by reduction of the rf drive amplitude. $AF^+$ ions (mass 410 amu) are ejected first; inset (a): extraction of a different sample of non-laser-cooled barium and $AF^+$ ions at ∼300 K; the small left-hand peak is due to SC $CO_2^+$ impurities; inset (b): extraction of a laser-cooled $Ba^+/AF^+$ ion cloud at a temperature of a few hundred mK (fluid state) [52].

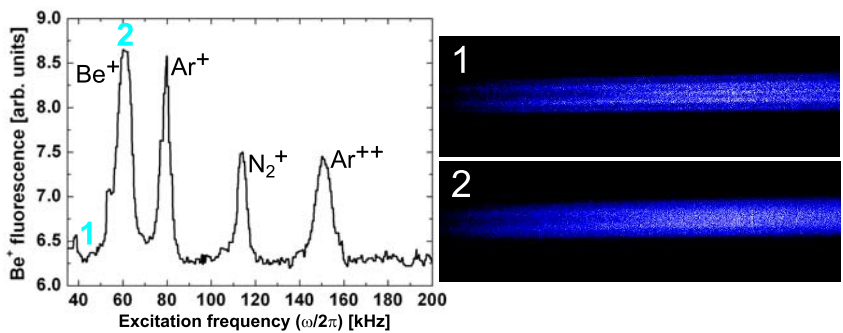

Figure 1.19: Motional frequency spectrum of a crystal containing $Be^+$, $Ar^+$, $N_2^+$, and $Ar^{2+}$ at 20 mK. Excitation of a particular (mass-specific) trap oscillation mode pumps energy into the motion of that species which also leads to heating of the atomic coolants. This increases the atomic fluorescence level which is simultaneously recorded using a photomultiplier tube (graph) and a CCD camera (images 1,2). Images were taken at the excitation frequencies indicated (1,2) in the spectrum. Excitation of a particular mode leads to blurring of the crystal structure, as observed with the camera (2), whereas when the excitation field is not resonant with a particular mode the structures are more pronounced (1); see also Fig. 1.22.

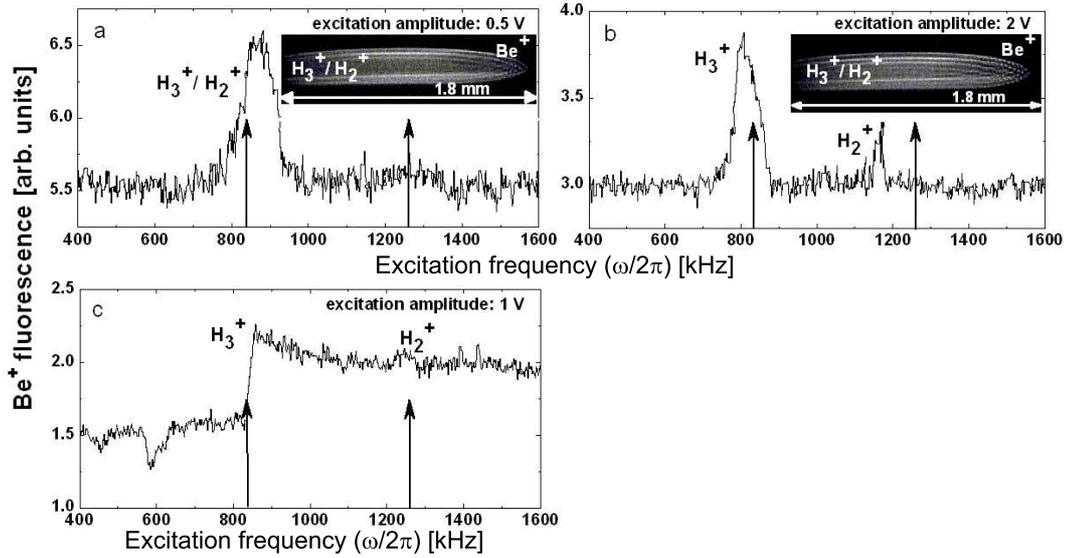

Figure 1.20: Interaction-induced strong (a,b) and weak (c) coupling between radial modes. Motional frequency spectrum of (a) an ion crystal (inset) containing $\approx 1400$ Be$^+$ and $\approx 1300$ SC ions ($\approx 700$ H$_2^+$ and H$_3^+$, and $\approx 600$ BeH$^+$) at $\approx 20$ mK. (b) ion crystal in (a) after partial removal of lighter SC ions (reduced size of the core), containing $\approx 1350$ Be$^+$ and $\approx 1200$ SC ions: $\approx 450$ H$_3^+$, $\approx 100$ H$_2^+$, and $\approx 650$ BeH$^+$. (c) Spectrum of an ion crystal containing Be$^+$, H$_2^+$, and H$_3^+$ ions in the gaseous/fluid state. Calculated single-particle secular frequencies (arrows): 840 kHz (H$_3^+$) and 1260 kHz (H$_2^+$). The feature at $\approx 580$ kHz is attributed to the second harmonic of the Be$^+$ radial mode (at 280 kHz) [70].

Fig. 1.21 shows examples of multi-species ion crystals and their (radial) motional resonance spectra. Fig. 1.21 b shows the spectrum of a cold beryllium ions crystal containing Be$^+$, H$_3^+$, H$_2^+$, and H$^+$ ions at 15 mK. Even though the measured motional frequencies are shifted compared to their calculated single-particle frequencies, particle identification is possible. Fig. 1.21 d shows a spectrum of a beryllium ion crystal containing C$_4$F$_8^+$ ions and various fragment ions.

Usually, the observed resonance frequencies are determined by a superposition of several, sometimes opposing line shifting effects. For example, the motional frequencies depend on the fractions of particles contained in the crystal. For strong coupling between the ion species, this can lead to a significant shift and broadening of individual features in the spectrum, so that they cannot be resolved experimentally (Fig. 1.20 a). However, even for weaker coupling (Fig. 1.20 b), various, sometimes more subtle, line shifting effects can be present, caused, e.g., by space charge effects, trap anisotropies, or the finite amplitude of the excitation field. The observed frequency positions also depend on the sweep direction of the excitation field and the plasma temperature. Finally, the state of the ion plasma, crystalline or gaseous/fluid, affects the motional resonances measured (see Fig. 1.20 c).

Therefore, an unambiguous identification of particles embedded in large, mixed-species ion crystals is often not possible based on the experimental measurements only. Comparison with MD simulations leads to an improved interpretation, finally enabling a more accurate interpretation of spectroscopic and chemical experiments.

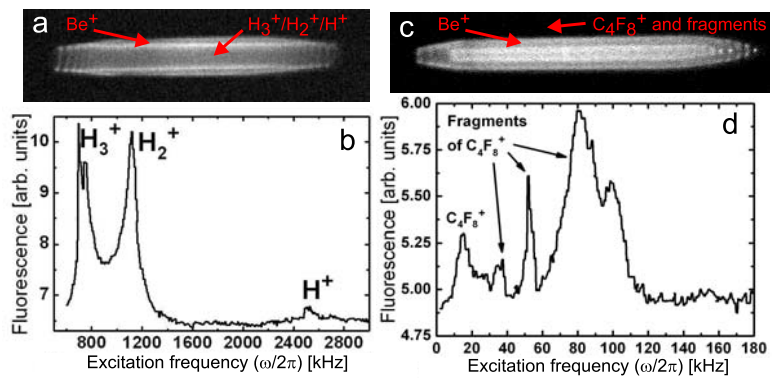

Figure 1.21: Motional frequency spectra of mixed-species crystals. (a) Crystal containing $Be^+$, $H_3^+$, $H_2^+$, and $H^+$ ions at $\approx 15$ mK. (b) Spectrum (low-mass range) of the crystal in (a). Measured frequencies are shifted compared to the calculated single-particle frequencies (840 kHz for $H_3^+$, 1260 kHz for $H_2^+$, and 2520 kHz for $H^+$; the $Be^+$ frequency is at 280 kHz), due to Coulomb coupling between LC and SC ions. (c) Crystal containing $Be^+$, $C_4F_8^+$, and various fragments of $C_4F_8^+$ at $\approx 20$ mK. (d) Spectrum (high-mass range) of the crystal in (c). The calculated single-particle frequency for $C_4F_8^+$ (13 kHz) agrees well with the measured value (15 kHz), due to smaller Coulomb coupling to all other species. The fragment ions act as conducting layer and ensure efficient sympathetic cooling of the $C_4F_8^+$ [52].

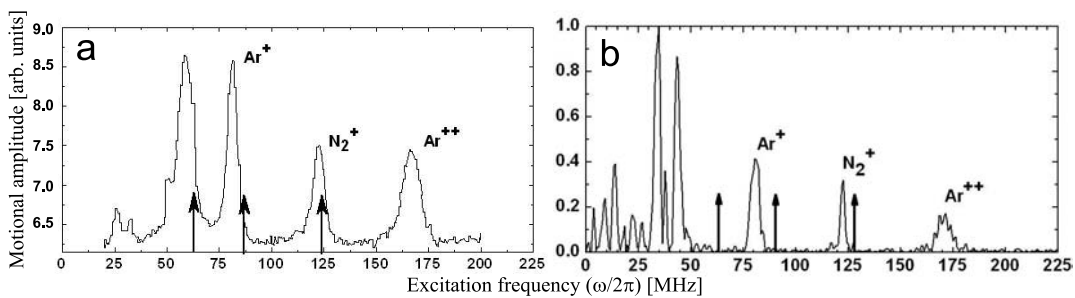

Figure 1.22: Measured (a), as in Fig. 1.19 a, and simulated (b) motional frequency spectrum of the cold (<20 mK) multi-species ion crystal containing $Be^+$, $N_2^+$, $Ar^+$, and $Ar^{2+}$ shown in Fig. 1.19 [70]. Arrows: single-particle frequencies.

As an example, in Fig. 1.22 the measured motional frequency spectrum for a mixed-species ion crystal is compared to the simulated spectrum. First, the number of ions is determined from a MD fit to the CCD images. Then the motional spectrum is obtained by starting with the equilibrium state of the ensemble, shifting the position of the SC particles radially and then evolving the system. The $x$ coordinate values are then Fourier transformed. The measured spectrum shows a fairly complicated structure with features at 58 kHz, 82 kHz, 122 kHz, and 166 kHz. The feature at 82 kHz is due to sympathetically crystallized $Ar^+$, whereas the features at 122 kHz and 166 kHz are attributed to $N_2^+$ and $Ar^{++}$ ions, respectively. The calculated single-particle secular frequencies for $Ar^+$, $N_2^+$, and $Ar^{++}$ are 63 kHz, 90 kHz, and 126 kHz. This shows how the Coulomb coupling between the ions produces significant shifts compared to the single-particle frequencies. The experimental feature at 58 kHz is not well reproduced, but can be attributed to the excitation of the axial $\omega_z$ $Be^+$ mode, which in the simulations appears as a double peak, due to small anisotropies of the trap potential.

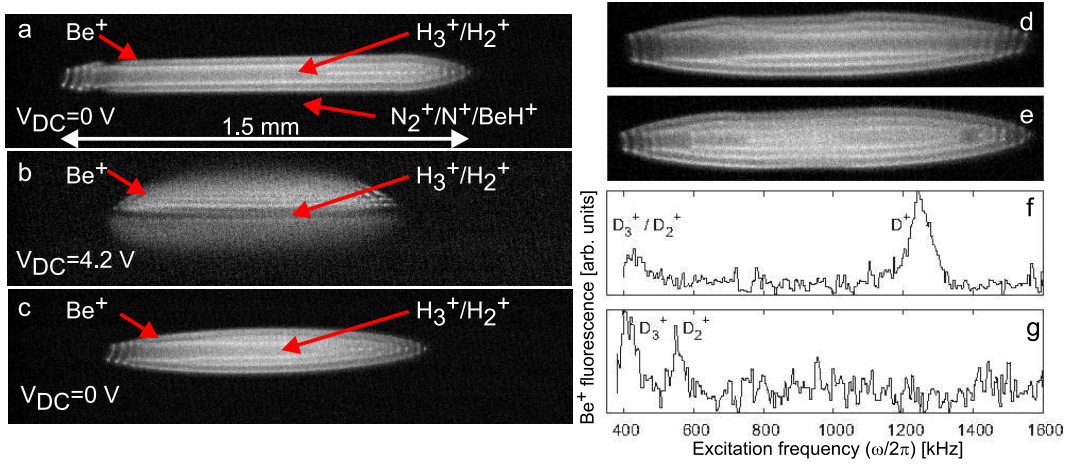

Figure 1.23: Left: Species-selective removal of heavy SC ions from an ion crystal, by applying a static quadrupole potential $V_{DC}$ to the trap electrodes. CCD images are taken before (a), during (b), and after (c) removal of the particles. After removal a re-ordering of the Be$^+$ shells is observed. The temperature of the crystal is slightly reduced, since heating effects for particles closer to the trap axis are smaller. Right: Removal of light SC ions. CCD images and spectra before (d,f) and after (e,g) the ejection of D$^+$ ions [47]. In (f), the individual contributions of D$_3^+$ and D$_2^+$ cannot be resolved, due to strong Coulomb coupling between the SC species (in contrast to (g)). The spectra were taken with different amplitudes of the excitation field, due to different ion numbers involved.

### 1.5.4 Species-selective ion removal

During loading and ionization of neutral gases in the trap along with the species under study various impurity ions are often produced by chemical reactions between LC or SC ions and the neutral gases. Such impurities can complicate precise measurements of motional resonances or systematic studies on pure or few-species ion plasmas. In particular, for the applications described in Sections 1.6 and 1.7, crystals with a single species of SC particles are required. It is therefore necessary to remove unwanted species from the crystal and leave all other species as intact as possible.

Ions with a mass-to-charge ratio *larger* than that of the atomic coolants are located outside the LC ion shells and can be removed from the trap selectively by adding a static quadrupole potential $V_{DC}$ to the electrodes. For appropriate strength, ion motion becomes unstable along one radial direction, leading to the ejection of that particular species from the trap [32, 66].

After switching off the static quadrupole potential again, the absence of the removed species causes a change of shape of the crystal. This procedure is illustrated in Fig. 1.23 (left), where N$_2^+$, N$^+$, and BeH$^+$ ions were removed from a cold ($\approx$20 mK) beryllium ion crystal. The crystal changed from approximately cylindrical to ellipsoidal. Note that the dark core of the crystal containing lighter SC ions (hydrogen molecular ions, H$_3^+$ and H$_2^+$) was not affected.

Particles with a mass-to-charge ratio *smaller* than that of the atomic coolants, and there-fore located closer to the trap axis, can be ejected from the crystal in a different way; see Fig. 1.23 (right). By detuning the cooling laser far from resonance, the ion crystal in Fig. 1.23 d undergoes a phase transition to a disordered (gaseous) state. In this situation, the coupling between different ion species is much weaker than in the crystalline state, and the secular motion of the unwanted species can be strongly excited, ejecting those ions from the trap, with almost negligible effect on other species. The remaining ions can then be re-crystallized by re-tuning the cooling laser back close to resonance (Fig. 1.23 e). This procedure can be repeated to remove as many different species as required. This and the previous procedure can be performed in sequence and thus pure two-component crystals can be obtained.

## 1.6 Chemical reactions and photofragmentation

### 1.6.1 Ion-neutral chemical reactions

Reactive and nonreactive collisions of ions with neutrals are of general interest in chemistry [71]. Ideally, the reactions would be studied as a function of collision energy, spanning the range from $\mu$eV to eV. Because of the experimental challenges, so far studies of ion-neutral reactions at low temperatures are still very few. For example, using multipole ion traps and cold He buffer gas cooling, reaction rates and branching ratios of various chemical reactions could be deduced [7, 11]. In a quadrupole ion trap, sympathetically cooled molecular ions were used, in order to study the reaction $H_3O^+ + NH_3 \rightarrow NH_4^+ + H_2O$ at temperatures of $\approx 10\,\text{K}$ [38]. The study of such reactions at even lower temperatures could improve the understanding of ion-neutral reactions occurring in interstellar clouds [8, 12, 72, 73].

Samples of sympathetically crystallized cold ions open up the possibility to investigate these processes with a good accuracy (because the ion density can be determined) and eventually with resolution of individual quantum states. A first, and simplest step, is the study of reactions with neutral gas at 300 K.

This situation implies collision energies (in the center-of-mass frame) below room temperature if the neutral particles are lighter than the ions. The study of this regime is useful in itself but also for preparing future work on reactions at ultralow energies, e.g. between cold trapped ions and ultracold neutral atomic or molecular gases. We note that nonreactive, but vibration-rotation-deactivating collisions using cold atomic gases might be useful for achieving *internal* cooling of translationally cold molecules. Eventually, a further extension will be collisions with quantum-state resolution. This implies the necessity to prepare the reactants not only with well-defined collision energy (inaccuracy smaller than rotational level spacing), but where the partners are in particular quantum states and the quantum states of the products are also detected.

Many ion-neutral reactions, such as

$$XY^+ + A \rightarrow XA^+ + Y \tag{1.6}$$

$$XY^+ + BC \rightarrow XYB^+ + C \tag{1.7}$$

are exothermic and proceed without an activation barrier. For such reactions, the Langevin theory predicts a temperature-independent rate coefficient, for suitably low temperatures

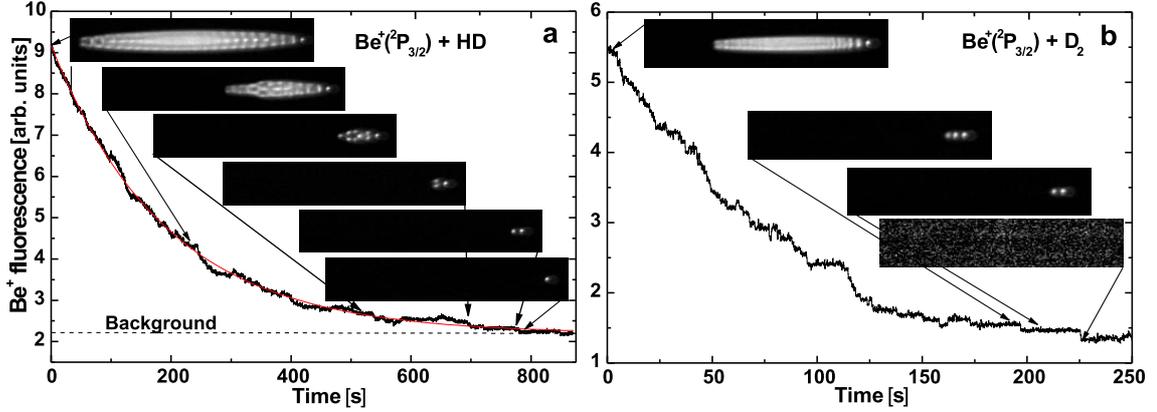

Figure 1.24: Chemical reaction between Be$^+$ and room-temperature molecular hydrogen gases. (a) Decay of a small Be$^+$ crystal ($\approx$160 Be$^+$ at $\approx$5 mK) after exposure to HD. The smooth line is an exponential fit to the data. The reaction coefficient is $k \approx 1.1 \cdot 10^{-9}$ cm$^3$/s. (b) Decay of a crystal with $\approx$45 Be$^+$ ions after exposure to D$_2$. Discrete steps in the fluorescence towards the end of the process are due to the formation of a single BeD$^+$ and of two BeD$^+$, respectively. Note how the right end of the crystal almost does not shift. The SC product ions cluster on the left side of the crystal because they do not experience the light pressure from the cooling laser, which propagates to the right [48].

[74, 75]:

$$k_L = Q\sqrt{\frac{\pi\alpha}{\epsilon_0 \mu}}, \qquad (1.8)$$

where $Q$ is the charge of the ion, $\alpha$ is the polarizability of the neutral reactant, and $\mu$ is the reduced mass of the particle pair.

**Reactions of laser-cooled atomic ions**

The cold ion-neutral reactions most easily studied are those involving the laser-cooled atomic ions. First examples studied were the formation of cold trapped CaO$^+$ ions by the reaction between laser-cooled Ca$^+$ ions and neutral O$_2$ [32, 76]. Using the CaO$^+$ ions formed, the back-reaction CaO$^+$ + CO $\to$ Ca$^+$ + CO$_2$ was observed [32]. Furthermore, the formation of cold trapped MgH$^+$ by the reaction between laser-cooled Mg$^+$ ions and neutral H$_2$ was also observed [30]. Reaction rates and branching ratios were deduced.

A reaction requiring photoactivation is between Be$^+$ and neutral molecular hydrogen gas, shown in Fig. 1.24 [48]. This reaction does not proceed with the beryllium ion in its ground electronic state. But when laser-cooled Be$^+$ ions are excited to the $^2$P$_{3/2}$ state, reaction occurs with a rate comparable to the Langevin rate:

$$(\text{Be}^+)^* + \text{HD} \to \text{BeH}^+ + \text{D} \qquad (1.9)$$

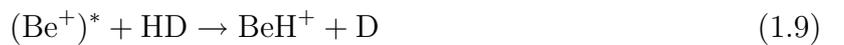

$$\to \text{BeD}^+ + \text{H}. \qquad (1.10)$$

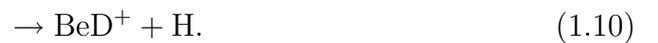

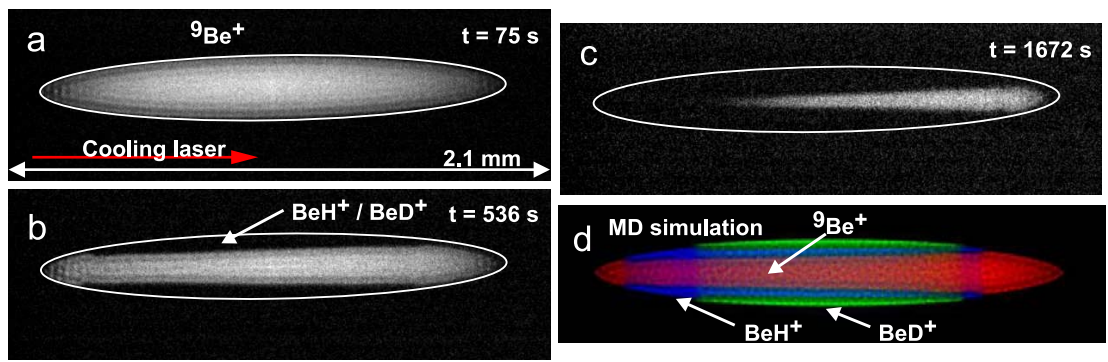

Figure 1.25: CCD camera image of an initially pure large $Be^+$ crystal following exposure to HD gas, leading to the formation of cold $BeH^+$ and $BeD^+$ via chemical reactions. The ellipse is a fit to the initial $Be^+$ crystal boundary. The molecular ions formed are located in the region enclosed by the ellipse. Ion numbers (determined via MD simulations) are: (a) 2100 $Be^+$, (b) 900 $Be^+$, 1200 $BeH^+$ and $BeD^+$ (in approx. equal amounts), (c) 150 $Be^+$, 1700 $BeH^+$ and $BeD^+$. (d) MD simulation of the crystal in (b), with 900 $Be^+$, 600 $BeH^+$, and 600 $BeD^+$ ions at approx. 15 mK.

The product ions were sympathetically crystallized. The reaction could be followed until the last few $Be^+$ ions reacted away, i.e. with a particle number resolution down to the single-particle level.

A reaction that does not rely on the atomic ion being laser-excited is the production of $^{138}BaO^+$ molecular ions via background $CO_2$ molecules [45]:

$$Ba^+ + CO_2 \rightarrow BaO^+ + CO \ . \tag{1.11}$$

Such reactions are relatively simple to characterize because (i) sympathetic cooling of the product ions permits their identification by mass spectrometry; (ii) reaction rates can be deduced from the time evolution of the laser-cooled ion number (observed using a CCD camera), or from the observation of the atomic fluorescence rate (using a PMT).

An interesting issue is the effectiveness with which the reaction product ions are sympathetically cooled. This can be seen in Fig. 1.25. The MD analysis indicates that nearly all product ions were sympathetically crystallized. A reason for this might be that the neutral product (H or D) is much lighter than the ion product and thus carries away most of the kinetic energy set free in the exothermic reaction.

**Reactions of molecular ions**

Perhaps the most fundamental ion-neutral reaction is the astrophysically important, exothermic chemical reaction of $H_2^+$ ions and neutral $H_2$ gas [72],

$$H_2^+ + H_2 \rightarrow H_3^+ + H \ , \tag{1.12}$$

which involves two of the most fundamental molecular ions [77].

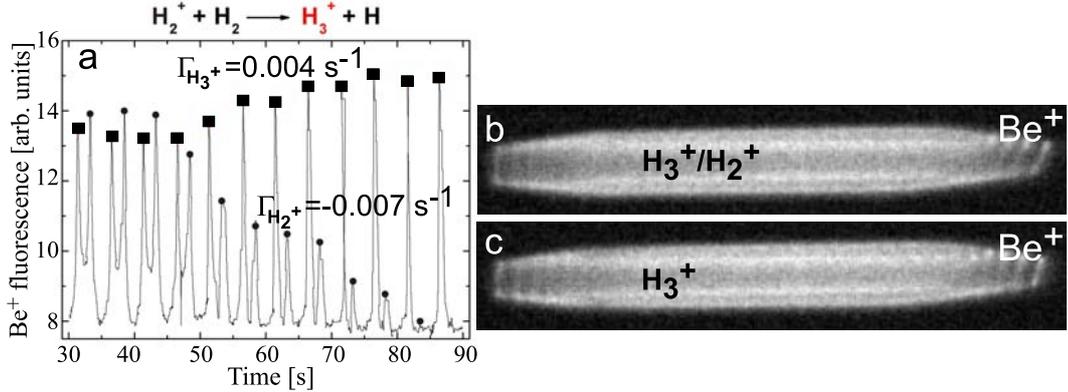

Figure 1.26: Chemical reaction between cold $H_2^+$ and room-temperature $H_2$. (a) Motional frequency spectrum of a cold multi-species ion crystal containing $Be^+$, $H_3^+$, and $H_2^+$ ions, during exposure to neutral $H_2$ gas. $H_2^+$ and $H_3^+$ resonances are denoted as circles and squares, respectively. CCD images taken before (b) and after (c) exposure to $H_2$. No significant loss of $Be^+$ ions occurred.

This reaction was studied after producing cold $H_2^+$ ions by electron impact ionization of neutral $H_2$ and sympathetic cooling using laser-cooled $Be^+$ ions and subsequent exposure to $H_2$ gas; see Fig. 1.26. Because the sympathetic cooling of the product ions is so efficient, the number of molecular ions does not change appreciably and almost no change in the Coulomb crystal structure occurs. The product ions can be identified via secular excitation mass spectrometry; see Fig. 1.26 a. The heights of the peaks in the secular spectrum are a measure of the ion numbers of SC ions in the crystal. During exposure of the cold ($\approx$10 mK) $H_2^+$ ions to room-temperature $H_2$ gas, in the motional resonance spectrum a decrease of the $H_2^+$ ion number is observed, while the $H_3^+$ ion number increases from an initially present amount. Fitting the peak maxima decay or increase yields $H_2^+$ loss and $H_3^+$ formation rates ($\Gamma$) for both molecular ion species. The difference between the two values found is probably due to nonlinear dependence of the signal on ion number due to mode coupling effects. We expect that a detailed understanding of the motional resonance spectrum could be obtained from MD simulations.

After cleaning of $BeH^+$ ions generated in addition to Eq. 1.12 using the procedure mentioned in Sec. 1.5.4, only $H_3^+$ and $Be^+$ ions remain; see Fig. 1.27 a. Thus, this is an example of the production of a sympathetically cooled molecular ion sample that contains only one molecular species.

The ability to produce chemical reactions with a high degree of accuracy and control was used to implement a fast and efficient method for molecule-to-molecule conversion. Cold $H_3^+$ ions were exposed to room-temperature HD gas, leading to the formation of $H_2D^+$ ions, a species of astrophysical interest [14], via the reaction

$$H_3^+ + HD \rightarrow H_2D^+ + H_2 , \qquad (1.13)$$

see Fig. 1.27 b. This reaction is exothermic by about 232 K [7]. After subsequent exposure to room-temperature $H_2$, the ions were converted back to $H_3^+$ via

$$H_2D^+ + H_2 \rightarrow H_3^+ + HD . \qquad (1.14)$$

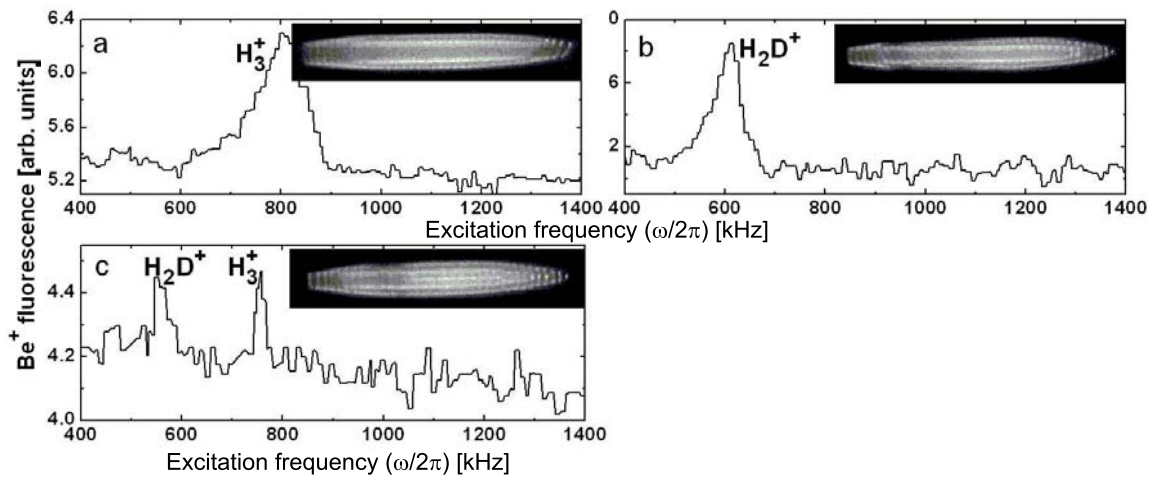

Figure 1.27: Molecule-to-molecule conversion using SC molecular ions. (a) Motional (secular) frequency spectrum of a Be$^+$ crystal containing cold H$_3^+$. (b) After exposure to HD gas, most H$_3^+$ ions are converted into H$_2$D$^+$. (c) Subsequent exposure to H$_2$ gas leads to partial back-conversion of H$_2$D$^+$ to H$_3^+$. According to the CCD images, loss of hydrogen molecular ions occurs, while Be$^+$ loss is negligible.

The barrier for this back-reaction is overcome by the thermal energy of the H$_2$ molecules or internal energy of the cold H$_2$D$^+$ ions. Depending on the exposure time to the neutral H$_2$ gas, back-conversion efficiencies close to unity can be reached. Loss of hydrogen molecular ions (seen as decrease in size of the dark crystal core) is mainly due to reactions with nitrogen molecules present in the background gas, leading to the formation of heavier SC ions embedded outside the Be$^+$ ensemble [49]. The presence of the reaction products is obvious from the increasingly flattened shape of the ion crystal in Fig. 1.27 (going from (a) to (b)).

Heteronuclear diatomic ions with large vibrational and rotational frequencies are promising systems for high-precision laser spectroscopy and fundamental studies, such as tests of time-independence of the electron-to-proton mass ratio. They can also serve as model systems for the implementation of schemes for internal state manipulation [78, 79]. Molecular hydrides, e.g., ArH$^+$ and ArD$^+$, are interesting examples, with the advantage of a relatively simple hyperfine structure of the ro-vibrational transitions [78, 80]. These hydrides were formed by the ion-neutral reactions [49] (Fig. 1.28 a-c)

$$\text{Ar}^+ + \text{H}_2 \rightarrow \text{ArH}^+ + \text{H}. \tag{1.15}$$

When the exposure to H$_2$ gas continues, the ArH$^+$ ions are converted to H$_3^+$,

$$\text{ArH}^+ + \text{H}_2 \rightarrow \text{H}_3^+ + \text{Ar}, \tag{1.16}$$

leading to a single-species molecular ion ensemble (Fig. 1.28 d,e).

The above is one of two possible reaction paths for the formation of cold H$_3^+$ ions; the second is

$$\text{Ar}^+ + \text{H}_2 \rightarrow \text{Ar} + \text{H}_2^+ \tag{1.17}$$

$$\Rightarrow \text{H}_2^+ + \text{H}_2 \rightarrow \text{H}_3^+ + \text{H}. \tag{1.18}$$

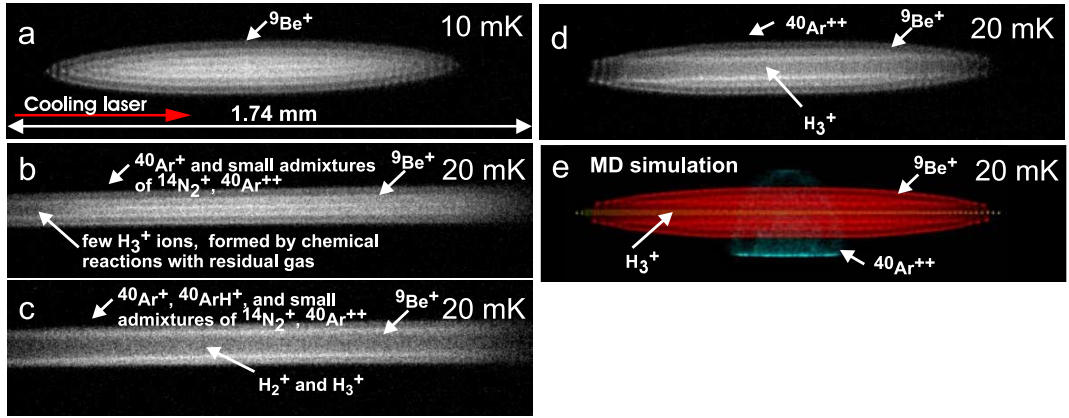

Figure 1.28: Production of cold molecular ions using sequential chemical reactions. (a) CCD image of a pure Be$^+$ ion crystal, (b) after loading with Ar$^+$ ions, (c) after H$_2$ inlet, leading to the formation of (mainly) H$_3^+$ ions and a smaller fraction of H$_2^+$ ions (as found by secular excitation mass spectroscopy), (d) after removal of Ar$^+$, ArH$^+$, and heavier contaminants, and full conversion of H$_2^+$ ions into H$_3^+$. Ar$^{++}$ ions were deliberately not removed. (e) Ion numbers and temperatures for the crystal in (d) are obtained from MD simulations: $\approx 1150$ Be$^+$ ions, $\approx 100$ H$_3^+$ ions, and $\approx 30$ Ar$^{++}$ ions at $\approx 20\,\mathrm{mK}$ [49].

For both paths (Eqs.(1.15) and (1.16) and Eqs.(1.17) and (1.18), respectively) all reactions are exothermic and are expected to proceed with a temperature independent Langevin reaction rate constant; see [49] and references therein for details.

The H$_3^+$ ion crystals produced in this way could be useful systems for exploring the chemistry of H$_3^+$. In particular, the study of state-specific reactions of H$_3^+$ via high-resolution infrared spectroscopy could provide valuable input for theories of ion-molecule gas-phase chemistry and precise calculations of molecular transition frequencies of this two-electron molecule.

A second example of the production of molecular ions via a sequence of two ion-neutral reactions with different neutral reactants is the production of HO$_2^+$. The first reaction is eq.(1.12), followed by

$$H_3^+ + O_2 \rightarrow HO_2^+ + H_2 .  \tag{1.19}$$

This reaction is nearly thermoneutral [81, 82]. Since the product ions are heavier than the atomic coolant ions, they crystallize on the outside of the atomic ion ensemble. The dark core of H$_3^+$ ions therefore decreases and the reaction can be directly followed from the CCD images, see Fig. 1.29 [48].

In summary:

- reactions can be produced in which the laser-cooled and crystallized atomic ions are "spectators" only, opening up the study of a large variety of reactions

- in particular, it is possible to study reactions between simple molecules (diatomics, hydrogen molecules), which are of importance for astrochemistry

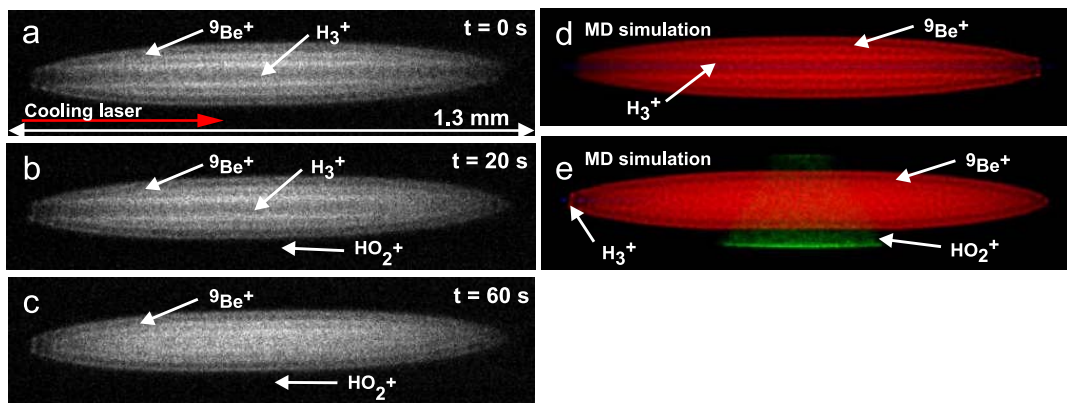

Figure 1.29: (a)-(c) Images of an initially pure Be$^+$-H$_3^+$ ion crystal exposed to neutral O$_2$ at $3 \times 10^{-10}$ mbar. H$_3^+$ disappears from the crystal core as HO$_2^+$ molecules are formed and embedded in the outer crystal region. Their presence leads to a slight flattening of the upper and lower edge of the crystal in (c). (d) MD simulation of the crystal in (a) containing 1275 Be$^+$ and 80 H$_3^+$ ions at $\approx 30$ mK. (e) MD simulation of the crystal in (c) containing 1275 Be$^+$, 3 H$_3^+$ and 75 HO$_2^+$ ions at $\approx 30$ mK.

- reactions can be studied on small ensembles; those are more likely to be prepared in well-defined quantum states in the future

- the rates of exothermic barrierless reactions can be set (by choice of neutral gas pressure) such that the reaction can be slow, permitting the application of nondestructive detection, and, in the future, laser manipulation of internal states of the molecules and atoms.

### 1.6.2 Photofragmentation of polyatomic molecules

Studies of laser-induced fragmentation of molecules are an important topic in chemical physics and can be useful, among others, for:

- the development of techniques to measure the internal state distribution of cold trapped molecules,

- the measurement and manipulation of branching ratios of dissociating channels in simple molecular systems,

- the development of theoretical photofragmentation models based on first principles in *ab initio* computational quantum chemistry, compare, e.g., [83],

- the study of photofragmentation and of conformational dynamics of complex polyatomic molecules, such as proteins and polymers, in the gas phase,

- as a (destructive) means to probe the change in population of vibrational and rotation levels upon ro-vibrational excitation [15].

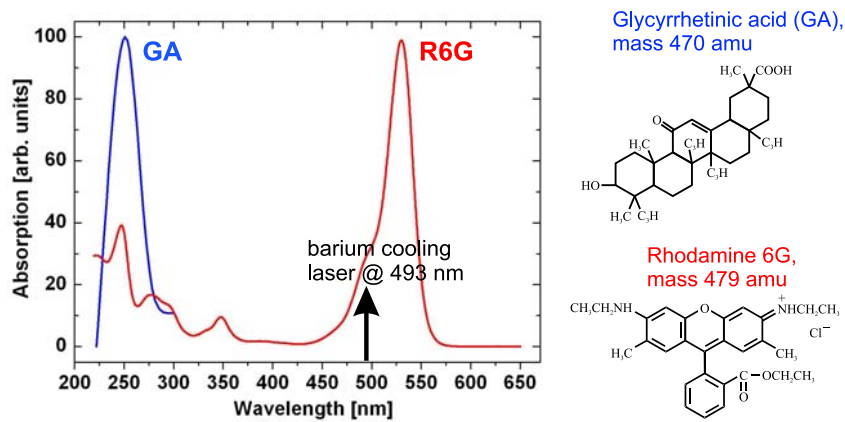

Figure 1.30: Absorption spectra of solvated GA and Rhodamine (R6G) ions and their structure.

In traps using buffer gas cooling photofragmentation is an established method [15], where it is applied to polyatomic ions [15, 16, 84, 85]. In connection with sympathetic cooling of MgH$^+$ molecular ions, two-photon dissociation was demonstrated and the branching ratio of the two possible dissociation channels, Mg + H$^+$ and Mg$^+$ + H, was investigated [76]. In all of these experiments, pulsed lasers were employed. Photodissociation of HD$^+$ is described in the following section.

One of the advantages of cold molecular ion ensembles is the long lifetime. Therefore photofragmentation studies can be performed with continuous-wave radiation. This appears as an interesting regime, since multiphoton processes can be avoided, and measured fragmentation rates can therefore be more easily compared with theoretical results.

The absorption spectra of Rhodamine 6G ions (R6G$^+$, mass 479 amu) and Glycerrhetinic acid ions (GAH$^+$, mass 470 amu) in solution are shown in Fig. 1.30. They indicate that R6G$^+$ will absorb cooling laser light and might fragment as consequence, while GAH$^+$ will not. This is indeed observed on the cold ions. Figs. 1.31 and 1.32 show photofragmentation of cold ($\approx$0.1 K) trapped R6G$^+$ and GAH$^+$ ions. Rhodamine 101 ions were also found to photodissociate in the presence of the cooling laser.

At least three different fragmentation detection techniques are possible: (i) observation of the change of shape of the cold ion crystal as fragments (lighter than the atomic coolant) are sympathetically cooled into the center of the crystal, (ii) extraction and counting of the parent ions and of the fragmentation products, (iii) motional resonance mass spectroscopy.

The first two methods are shown in Figs. 1.31 and 1.32. The third method is described in detail in [36]. This latter method is important because the first two methods have inherent limitations: for certain systems, the interpretation of the structures in the first method appears too difficult, because of the large number of different product species. The second method is destructive and therefore systematic measurements, e.g. as a function of fragmentation laser intensity, will be very time consuming. For such systems, the third method can be an alternative.

A particular advantage of these techniques are the long storage times of many minutes (up to hours) in the well-defined and nearly collisionless environment of an ion trap in an ultrahigh vacuum chamber. This allows for the investigation of slow destruction processes such as the photodissociation of large biomolecules. The methods could also be applied for a (systematic) study of highly resolved photodissociation spectra using low intensity, tunable continuous-wave lasers with narrow linewidths. In order to reduce the spectral congestion due to the presence of different conformers, a cooling of the internal degrees of freedom would be advantageous. This could, for example, be implemented by radiative cooling in a cryogenic environment [36].

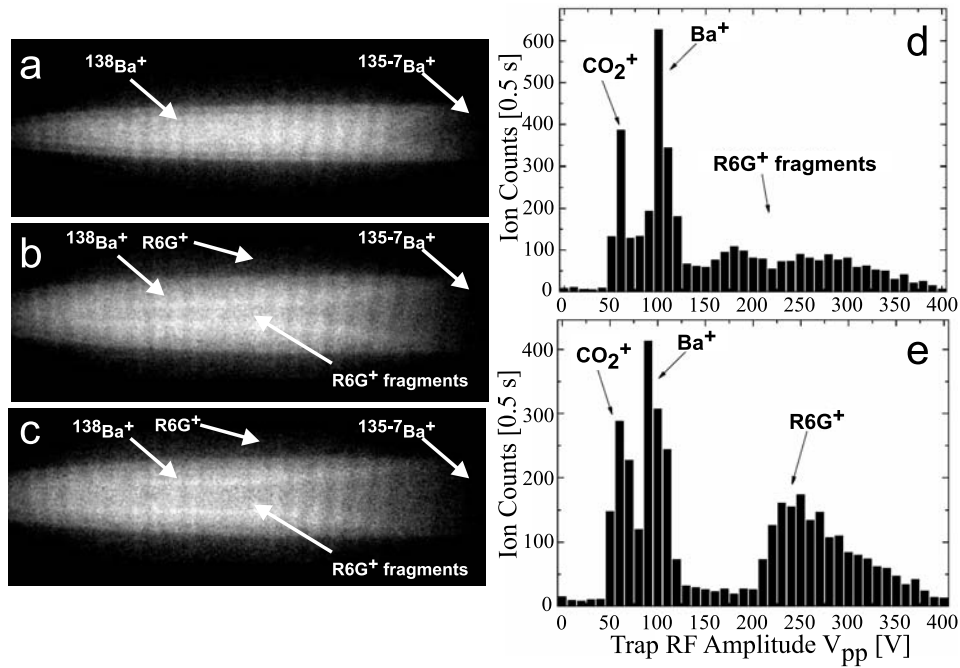

Figure 1.31: Photodissociation of cold, singly charged (protonated) Rhodamine 6G ions by the 493 nm $Ba^+$ cooling laser. Left: Cold $Ba^+$ ion crystal before (a) and after (b) loading of $R6G^+$. During loading the $R6G^+$ ions (located outside the $Ba^+$ ion ensemble) are exposed to 493 nm-radiation and $R6G^+$ fragments are produced. Fragments lighter than $Ba^+$ are embedded inside the $Ba^+$ ion ensemble and lead to the dark core in (b). Fragments heavier than $Ba^+$ are located in between the $Ba^+$ and $R6G^+$ ion ensembles and can also be dissociated by the cooling light. Typically, ions heavier than $Ba^+$ lead to a deformation of shape of the $Ba^+$ ion crystal (depending in size on their number and mass-to-charge ratio) which is, however, not noticeable here. (c) CCD image taken after 60 s of exposure of the $R6G^+$ to the $Ba^+$ cooling laser (493 nm). The number of fragments lighter than $Ba^+$ and, thus, the size of the dark core have increased. Right: (d) Mass spectrum of an ion crystal similar to the one in (c) containing cold $Ba^+$, $CO_2^+$ (impurities; narrow left-hand peak), and various $R6G^+$ fragments heavier than $Ba^+$ (broad right-hand peak). $R6G^+$ fragments lighter than $Ba^+$ are probably also contained in the crystal, hidden in the spectrum under the pronounced $CO_2^+$ peak. The spectrum was obtained by extraction and counting of the ions. (e) Mass spectrum of an ion ensemble which was not laser-cooled and where $R6G^+$ fragments were therefore not formed. The peaks in the spectrum in (e) are broader than the (corresponding) peaks in (d), indicating translational temperatures for $Ba^+$, $CO_2^+$, and $R6G^+$ ions above 300 K. [36]

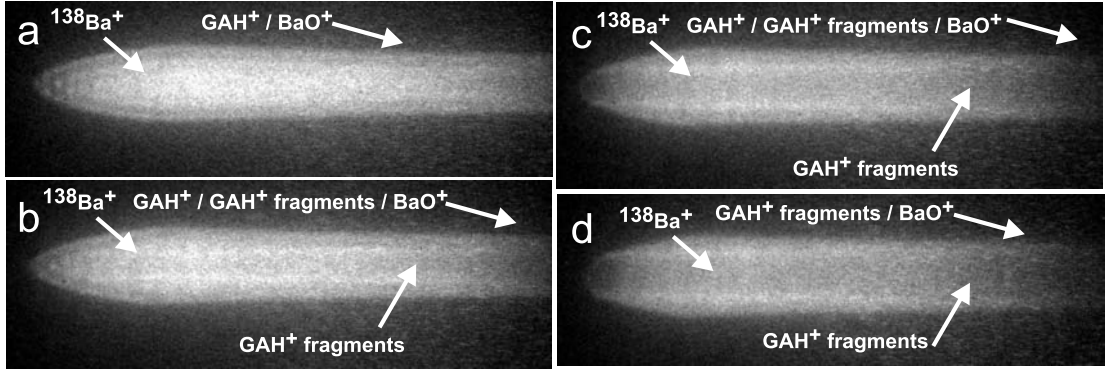

Figure 1.32: Photodissociation of cold, singly charged (protonated) Glycyrrhetinic (GA) acid ions by cw 266 nm UV radiation. Laser-induced fragmentation of the cold ($\approx$100 mK) GAH$^+$ ions leads to the appearance of a dark crystal core containing GAH$^+$ fragments. In addition, GAH$^+$ fragments heavier than Ba$^+$ have probably been formed. Images were taken before (a), during (b,c) and after (d) exposure to the UV radiation for several minutes; see also [36].

## 1.7 Rovibrational spectroscopy of molecular ions

### 1.7.1 Rovibrational spectroscopy

Traditionally measurements on molecular ions were performed in discharges [86], in ion beams [87] or in traps equipped with buffer gas cooling (>10 K) [13]. For SC trapped molecular ions, one wide-ranging application is rovibrational spectroscopy with significantly enhanced resolution and accuracy. This is enabled by the strong suppression of usual line-shifting and broadening effects due to collisions, high thermal velocities, and finite transit time.

Spectroscopy of rovibrational transitions in the electronic ground state can take advantage of these special conditions. The low translational temperature of the molecules increases the absorption rate significantly, and efficient excitation is possible even on weak overtone transitions [88]. This allows one to use simpler laser sources, which also simplifies experimentation. Due to the relatively long lifetime of vibrational levels of molecules ($\sim$ms to days), the potential line resolution can be huge. Spectroscopy of vibrational transitions on cold and localised samples of molecular ions was first achieved using HD$^+$ ions [89].

One interesting aspect is the dependence of vibrational and rotational transition frequencies on the ratio of electron mass to the mass of the nuclei [80, 90]. In the simplest case, a diatomic molecule, the fundamental vibrational and rotational transition frequencies scale approximately as:

$$v_{vib} \sim \sqrt{m_e/\mu}\, R_\infty, \ v_{ro} \sim m_e/\mu\, R_\infty. \tag{1.20}$$

Here, $\mu$ is the reduced mass of the two nuclei and $R_\infty$ is the Rydberg energy. These dependencies offer two opportunities: (i) the determination of the ratios $m_e/m_p$, $m_e/m_d$, $m_e/m_t$m, $m_p/m_d$, $m_p/m_t$ ($m_t$ is the mass of the tritium nucleus) by measurement of transition frequencies of the one-electron hydrogen molecular ions H$_2^+$, D$_2^+$, HD$^+$, HT$^+$ combined with high-precision ab-initio theory, and (ii) the search for a time-dependence of the ratio of elec-tron to nuclear mass [3]. The latter option is not restricted to diatomics and therefore offers the option of choice of molecular systems with suitably low systematic shifts.

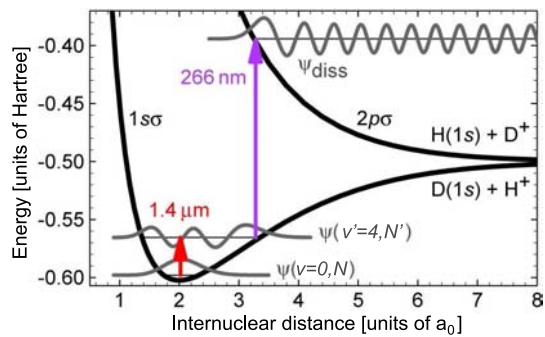

Figure 1.33: Principle of (1+1') REMPD spectroscopy of HD$^+$ ions. A tunable IR diode laser excites a rovibrational overtone transition. The excited HD$^+$ ions are dissociated using a cw 266 nm laser: HD$^+$(v'=4) + h$\nu$ → H + D$^+$ or H$^+$ + D [89].

Furthermore, it has been suggested that a possible variation of $m_e/m_p$ (and of the quark masses and the strong interaction constant) over time might be larger than a possible time variation of the fine structure constant $\alpha$ [91, 92, 93]; see also [94, 95, 96]. Recently, an indication of a variation of $m_e/m_p$ over billion-year timescales was reported in [97]. This was based on the comparison of laboratory measurements of Lyman bands of neutral H$_2$ with H$_2$ spectral lines observed in quasars, indicating that $m_e/m_p$ could have decreased over the near age of the universe.

Currently, the above constants are determined by Penning ion trap mass spectrometry and spin resonance. Their relative accuracies are $2 \times 10^{-10}$ for $m_p/m_d$ [98], $2 \times 10^{-10}$ for $m_p/m_t$, and $4.6 \times 10^{-10}$ for $m_e/m_p$ [98, 99], respectively. *Ab initio* calculations of energy levels in molecular hydrogen ions are approaching the limits set by the uncertainties in the values of those fundamental constants entering the calculations, the largest contribution originating from the uncertainty in the value for the electron-to-proton mass ratio $m_e/m_p$. Therefore, combining molecular hydrogen ion spectroscopy and theory has potential to eventually yield improved values for the mass ratios.

Rovibrational spectroscopy via fluorescence detection will not be feasible for a large class of molecular ions. In particular, it is not feasible in HD$^+$, since there is no stable excited electronic state. Therefore the fluorescence would be between vibrational levels. The corresponding low fluorescence rates would require sophisticated photon counting systems in the mid to far infrared, which is not practical. In such cases, a destructive technique can be used instead. In the simplest case, a (1+1') resonance enhanced multiphoton dissociation (REMPD) can be applied. The molecules are excited by an infrared (IR) laser and then selectively photodissociated from the upper vibrational state by a second, fixed-wavelength ultraviolet (UV) laser; see Fig. 1.33. The remaining number of molecular ions is measured as a function of the frequency of the excitation laser, see Fig. 1.34.

The molecular samples used are small (typically 40-100 ions). The spectroscopy requires the spectra to be obtained by repeated molecular ion production and interrogation cycles. The loss of HD$^+$ ions not only depends on the REMPD process, but also on transitions induced by blackbody radiation (BBR); see Fig. 1.35. The respective rates are below 1/s, but their effects are clearly seen in the experiment: they are responsible for the dissociation of all HD$^+$ ions, even though the laser only excites ions from a particular rotational level. For a comprehensive description of the process, the loss of HD$^+$ ions was modelled by solving the rate equations for the populations of all relevant $(v, J)$ levels including interaction with the IR and UV lasers and the BBR radiation at 300 K; see Fig. 1.36. The rate equation model reveals two different timescales at which the HD$^+$ number declines: a first, fast ($<$1 s) decay occurs when the IR laser selectively excites ions from a specific rovibrational level (here $v = 0, J = 2$) which are subsequently photodissociated. The magnitude of this decay depends on the laser intensities (it is small in the case of Fig. 1.35). A fraction of the excited ions cascade back to the ground vibrational state, but into other rotational levels, $v = 0, J \neq 2$. The lower level ($v = 0, J = 2$) is repopulated by BBR and by spontaneous emission at a slower rate and thus the molecular ions are then dissociated at a slower rate [100]. The spontaneous decay rates between vibrational levels differing by $\Delta v = 1$ are approx. 100/s and thus the vibrational relaxation is very fast compared to the rotational dynamics in $v = 0$.

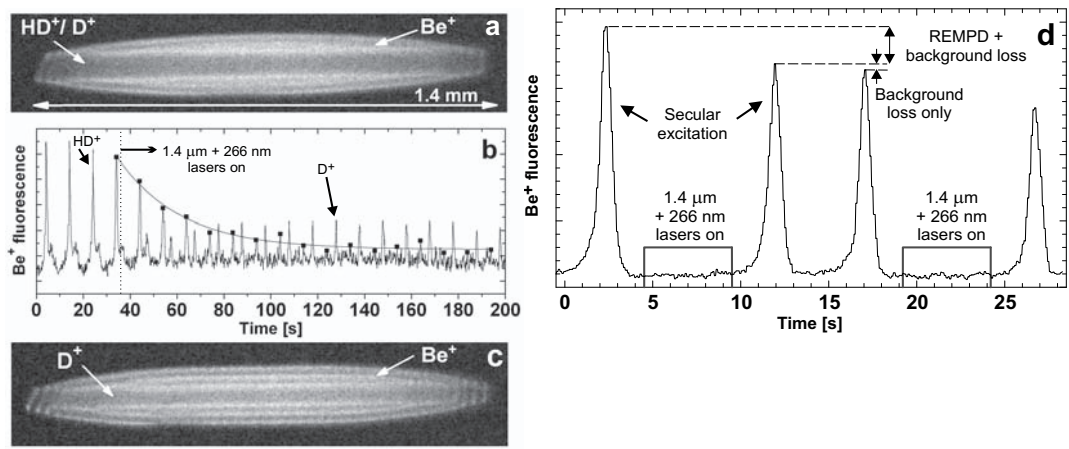

Figure 1.34: (1+1') REMPD spectroscopy of HD$^+$ ions. (a) Initial crystal: $\approx$1100 Be$^+$, $\approx$100 HD$^+$, and $\approx$20 D$^+$ at $\approx$20 mK. (b) Repeated secular excitation of the crystal in (a). The excitation frequency was swept between 500 and 1500 kHz every 4 s. The IR laser was tuned to the maximum of the $(v'=4, J'=1) \leftarrow (v=0, J=2)$ line. The curve is an exponential fit with a decay constant of 0.04 s$^{-1}$. (c) Crystal after dissociation of all HD$^+$: $\approx$1100 Be$^+$ and $\approx$50 D$^+$ at $\approx$20 mK. (d) Measurement cycle consisting of repeated probing of the HD$^+$ number before and after exposure to the spectroscopy lasers [89].

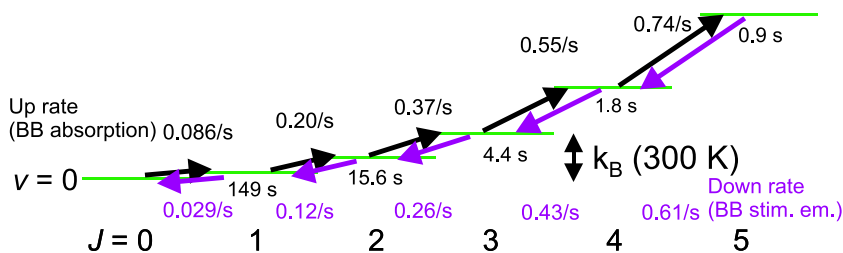

Figure 1.35: Blackbody(BB)-induced and spontaneous emission rates within the $v = 0$ rotational manifold of HD$^+$. Time values are the natural lifetimes. Rate values are for absorption of BB radiation and stimulated emission by BB radiation. BB temperature is 300 K.

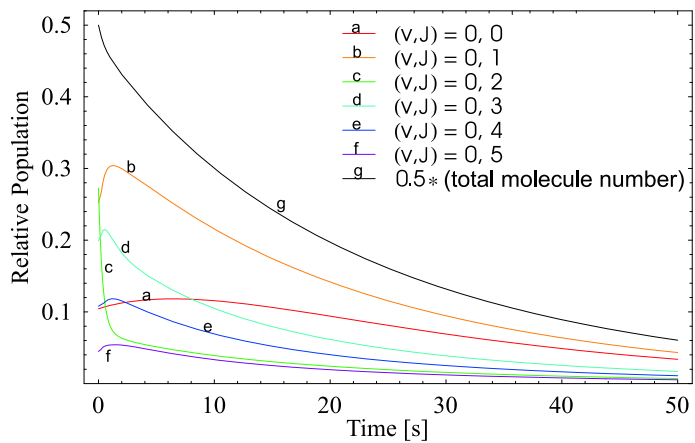

Figure 1.36: Rotational population dynamics for a translationally cold (20 mK), internally warm (300 K) collision-free ensemble of HD$^+$ ions during REMPD. The IR laser is set to the maximum of the $(v' = 4, J' = 1) \leftarrow (v = 0, J = 2)$ transition (1430 nm). The intensities of the lasers are $I(\nu_{IR}) = 0.32\,\text{W/cm}^2$, $I(266\,\text{nm}) = 0.57\,\text{W/cm}^2$. $J$ is the rotational quantum number. Hyperfine structure is neglected. At an intensity $I(266\,\text{nm}) = 10\,\text{W/cm}^2$ the time at which the ion number has decreased to 50% is reduced to 3.5 s.

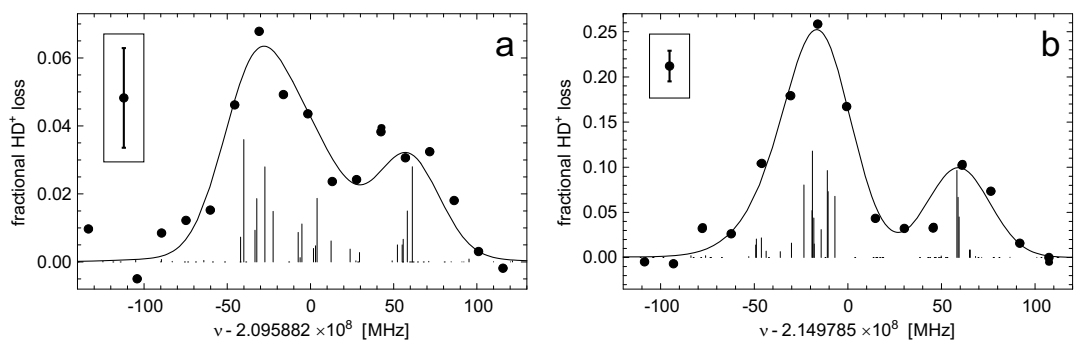

Figure 1.37: (a) The $(v' = 4, J' = 1) \leftarrow (v = 0, J = 2)$ transition at 1430 nm, (b) The $(v' = 4, J' = 3) \leftarrow (v = 0, J = 2)$ transition at 1395 nm. The curves are fits to the data (●), where the theoretical "stick" spectra were broadened by ≈40 MHz. The ordinate values are the molecular ion dissociation probabilities for a 5 s irradiation of 0.65 W/cm$^2$ IR and 10 W/cm$^2$ UV light. The insets show typical error bars [89].

Due to the weak coupling between external and internal degrees of freedom, the internal (rotational and vibrational) temperature of the HD$^+$ ions (see Sec. 1.7.2) is at approx. 300 K, in thermal equilibrium with the vacuum chamber, with a significant (>5%) population for rotational levels up to $J = 6$. Indeed, 12 transitions between 1391 nm and 1471 nm, from lower rotational levels $J = 0$ to $J = 6$ were observed using diode laser spectroscopy. A telecom-type diode laser with a linewidth of ∼5 MHz on a one-second timescale was used, and its frequency was calibrated with an accuracy of 40 MHz by absorption spectroscopy in a water vapor cell.

Detailed measurements for two rovibrational transitions, $(v' = 4, J' = 1) \leftarrow (v = 0, J = 2)$ and $(v' = 4, J' = 3) \leftarrow (v = 0, J = 2)$, are shown in Fig. 1.37. The measured data show par-tial resolution of the complicated hyperfine spectrum due to the coupling of four angular momenta (spins of electron, proton, deuteron, and rotational angular momentum). A linewidth of ≈40 MHz [89] was found and was attributed to Doppler broadening, mainly due to a parasitic axial micromotion of the ions.

### 1.7.2 Molecular thermometry

One application of rovibrational spectroscopy is molecular thermometry. This type of thermometry is a well-known diagnostic tool and used e.g. in combustion studies, where the rotational and vibrational temperatures are measured, often with coherent Raman spectroscopy, in order to determine their temperature or that of the surrounding medium.

An important question is to what extent the rotational motion of sympathetically cooled molecules couples to their translational motion, i.e., whether the internal state population can be manipulated by cooling the external degrees of freedom of molecules.

The internal temperature is expected to reach a stationary value due to the competition between the (long-range) collisions between charged particles, interaction with black-body radiation, and collisions with residual gas molecules. The cross section for collisions between a molecular ion and another charged particle (which in the ensembles discussed here can be another molecular ion or a laser-cooled atomic ion) that induce transitions between the rotational or vibrational states has been discussed in [101]. The transition probability between two (rovibrational) states $n$ and $n'$ is modelled by:

$$P(n \to n') = 4\pi^2 |\langle n'|m(R)|n\rangle|^2 |E_\omega(p)_{\omega=\omega_{nn'}}|^2, \qquad (1.21)$$

where $\langle n'|m(R)|n\rangle$ and $\omega_{nn'}(=|E_{n'} - E_n|)$ are the matrix element of the electric dipole moment $m(R)$ and the energy difference between the initial and the final states, respectively. $R$ is the internuclear distance for the molecular ion and $E_\omega(p)$ is the Fourier component of the electric field strength produced at the molecular ion by the incident charge with a given impact parameter $p$. In this model, the ion-ion interaction leads to internal heating, similar to the effect of the BBR.

At low relative collision energies (large $p$) in the cold ensembles, the electric field remains small, since the particles do not approach each other more than a few $\mu$m, and the rate of field strength change is small. The cross section and excitation/deexcitation probability drop with a high power of the relative energy, and are negligible compared to other effects. Similarly, the influence of trap or noise fields on the rotational distribution is expected to be negligible.

In an experiment on sympathetically cooled MgH$^+$ [102] data from rotational REMPD was compared with results from theoretical simulations. From the measurements it was concluded that the rotational temperature of the MgH$^+$ ions was higher than 120 K. However, the technique was not suited for measuring the rotational temperature of the ions accurately.

With the rotational-state resolved REMPD spectroscopy the rotational distribution can be measured directly. A study was performed on an ensemble of HD$^+$ at 10 mK translational temperature [103]. The HD$^+$ loss rate was measured for rovibrational transitions starting from lower rotational levels $J = 0$ to $J = 6$. For this purpose, the IR laser interrogating the ions was tuned to the maximum of individual rovibrational transitions and broadened to approximately 200 MHz, in order to cover their entire hyperfine spectrum.

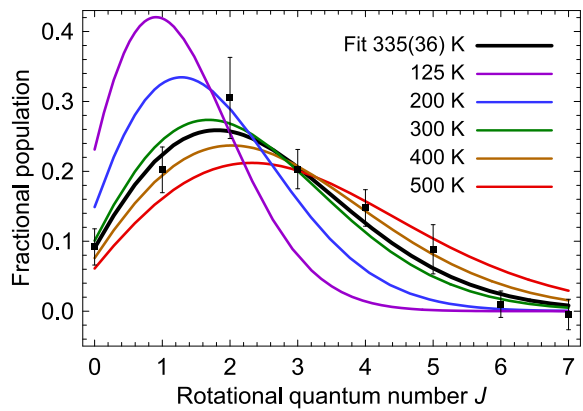

Figure 1.38: Rotational distribution of cold trapped HD$^+$ ions [103].

The fractional state population was deduced by fitting the data taking into account population redistribution due to coupling to the BBR. Under the experimental conditions given, the internal (rotational) degrees of freedom were found to be independent of the translational degrees of freedom. The effective rotational temperature found was close to room temperature (335 K), within the 11% measurement accuracy; see Fig. 1.38.

The method used is quite general and can be applied to other molecular species. Furthermore, in the near absence of background gas collisions it allows one to directly relate the rotational temperature of the ions to the temperature of the ambient blackbody radiation. This feature (among others, see [100, 104]) suggests the use of molecular ions, such as HD$^+$ or CO$^+$, for BBR thermometry with possible applications in frequency metrology, i.e., it may help to improve the accuracy of frequency standards based on trapped ions [103].

One far-reaching perspective is to perform experiments on translationally localized molecules prepared in specific quantum states. Several methods were proposed to cool the internal degrees of freedom of the molecules, using lasers or lamps [79], cryogenic techniques, or collisions with cold ($\mu$K-regime) neutral atoms. The internal temperature measurement method described could be a useful tool for future studies in this direction and for investigations of collisions in general.

### 1.7.3 High-resolution spectroscopy of molecular hydrogen ions

Among all molecular ions, the hydrogen ions deserve particular attention [90]. Molecular hydrogen ions are the simplest molecules in nature, containing only two nuclei and a single electron. Therefore, they have played an important role in the quantum theory of molecules since the birth of the field of molecular physics. More than 800 publications (mostly theoretical) have been written on these molecules over the last 35 years [105].

Molecular hydrogen ions are of fundamental importance in various aspects:

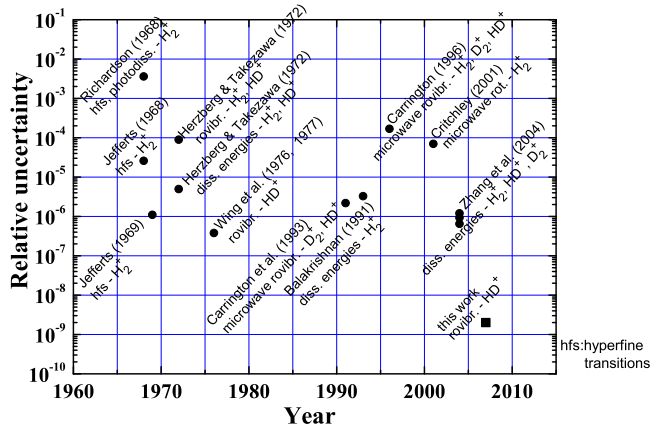

Figure 1.39: Experimental uncertainties for various energies of $H_2^+$, $HD^+$, and $D_2^+$.

- as benchmark systems for the test of advanced *ab initio* molecular calculations (in particular, QED contributions),
- for the measurement of $m_e/m_p$, $m_p/m_d$, and $m_p/m_t$,
- for the measurement of the deuteron quadrupole moment [113],
- for tests of Lorentz Invariance and time invariance of fundamental constants [3, 114],
- for tests of concepts for the internal state manipulation of molecules,
- for blackbody thermometry [103],
- for the study of collision processes and ion-neutral reactions,
- as model systems for the investigation of radiation-molecule reactions.

Several spectroscopic investigations on trapped room-temperature molecular hydrogen ions and on ion beams were performed in the past; see Fig. 1.39 [106, 87, 107, 108]. The lowest spectroscopic relative uncertainties reported so far were achieved in the experiments of Jefferts and of Wing et al. [87, 109], $\simeq 1 \times 10^{-6}$ and $4 \times 10^{-7}$ (in relative units), respectively. Dissociation energies were obtained with inaccuracy as low as $\simeq 6.5 \times 10^{-7}$ [110]. Over the years the theoretical accuracy of the rovibrational energies has steadily increased. They have recently been calculated by V. Korobov *ab initio* with an relative uncertainty below $1 \times 10^{-9}$ (70 kHz), including QED contributions [111]. Improved hyperfine structure calculations were also reported recently [112], with an estimated uncertainty of about 50 kHz. The different contributions to the calculated transition frequency for a particular rovibrational transition in $HD^+$ are displayed in Fig. 1.40.

### 1.7.4 Infrared spectroscopy of $HD^+$ ions with sub-MHz accuracy

Based on the REMPD spectroscopic method developed for cold molecular ions, the transition frequency for the rovibrational transition $(v' = 4, J' = 3) \leftarrow (v = 0, J = 2)$ at 1395 nm was measured using a novel narrowband grating-enhanced diode laser with resonant optical feedback locked to a femtosecond frequency synthesizer [115]. The latter was stabilized to a maser referenced to the Global Positioning System for longterm stability [116].

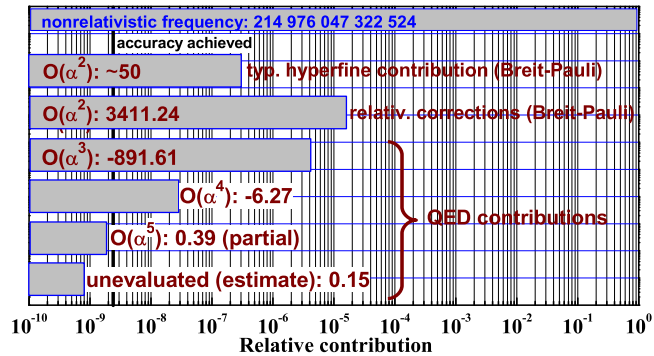

Figure 1.40: Contributions to the $(v' = 4, J' = 3) \leftarrow (v = 0, J = 2)$ transition frequency in HD$^+$ (in MHz) [111].

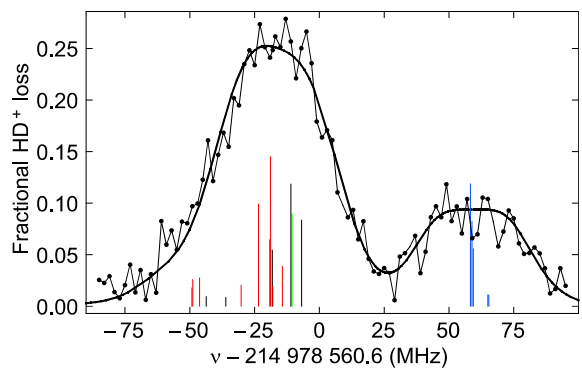

Figure 1.41: Spectrum of the $(v' = 4, J' = 3) \leftarrow (v = 0, J = 2)$ transition in HD$^+$ obtained using a precise laser spectrometer. The offset frequency in the abscissa label is the measured result for the unperturbed transition frequency and is accurate to within 0.5 MHz. The smooth curve is a fit to the data. The strongest transitions in the theoretical hyperfine line spectrum (underneath the fit) can be grouped by their spin configurations (see [115] for details).

The spectrum (Fig. 1.41) exhibits two peaks of about 40 MHz width. A major contribution to this is excess micromotion along the trap axis. The data allowed a fit to the theoretical hyperfine spectrum with an uncertainty of 0.45 MHz, limited by measurement noise. Consideration of systematic effects led to a slightly increased uncertainty of 0.50 MHz in the determination of the unperturbed (i.e. excluding hyperfine contributions) ro-vibrational transition frequency. The relative accuracy of 2.3 ppb is 165 times more accurate than the best previous results. The value for the unperturbed transition frequency deduced agrees with the calculated value from V.I. Korobov [111] (theoretical uncertainty 0.3 ppb) to within measurement accuracy. From the measured value the electron-to-proton mass ratio $m_e/m_p$ could be deduced with an accuracy of 5 ppb. The value is in good agreement with the 2002 CODATA value that has a relative uncertainty of 0.46 ppb [98].

It thus appears that the experimental method demonstrated for HD$^+$ provides a new approach for determination of the electron-to-proton mass ratio $m_e/m_p$ with improved accuracy in the near future. If the experimental accuracy improves by more than one order of magnitude and the uncertainty in the theoretical rovibrational energies improves by a factor of two, a competitive value for $m_e/m_p$ could be obtained. On the experimental side, this requires a trap and a spectroscopy laser that enables Doppler-free spectroscopy.

## 1.8 Summary and outlook

In summary, we recapitulate a few salient results obtained on sympathetically cooled molecular ions. It is possible to produce a variety of cold molecular ions, with masses between 2 and 12,400 amu at present. This range can be cooled with just two atomic coolant species. Experimental techniques for destructive and non-destructive characterization and analysis of the content of sympathetically cooled ensembles are available. Molecular dynamics simulations have proven useful to deduce properties of the ensembles. Chemical reactions between cold trapped atomic and molecular ions and neutral gases can be observed and used to determine reaction rates or to produce a variety of molecular ions in situ. At least for simple molecules, high-resolution rovibrational spectroscopy is feasible even if the internal temperature of the ions is at room temperature. A high spectroscopic accuracy can be achieved; as an example, the frequency of a rovibrational transition in cold, trapped HD$^+$ ions was measured with a relative uncertainty of $2.3 \times 10^{-9}$.

Based on the results obtained so far in the field of cold molecular ions, it appears both important and interesting to extend the toolbox of methods further. Some of the goals are

- extend sympathetic cooling to molecular ions with large masses, e.g. proteins
- develop nondestructive spectroscopic methods (quantum-logic-enabled spectroscopy, state-selective optical dipole force) both for simple and complex molecules
- extend high-resolution vibrational spectroscopy to other molecular ions of special interest, e.g. $H_3^+$, HeH$^+$, $H_2^+$
- push the accuracy and resolution of rovibrational spectroscopy further by achieving the Lamb-Dicke regime
- identify, by theoretical and experimental studies, molecular ions with systematic shifts so low that precise tests of fundamental laws can be performed
- demonstrate radiofrequency spectroscopy and two-photon spectroscopy
- develop practical methods for internal cooling (e.g. cold shields, optical pumping)
- use molecular ions as targets for studies of ion-neutral interactions (elastic and inelastic collisions, charge transfer, rotational and vibrational deactivation) at low temperature
- develop methods to study slow processes in polyatomic molecules, e.g. triplet-singlet decay rates, taking advantage of the near-collision-free environment and/or long observation times.

It is felt that significant progress can be made towards these goals in the near-future, so that the challenging scientific topics outlined in the introduction can be addressed.

*Acknowledgements* It is a pleasure to acknowledge the colleagues who in the course of the years contributed to the results described here, in particular D. Offenberg, C.B. Zhang, A. Ostendorf, U. Fröhlich, A. Wilson, P. Blythe, J. Koelemeij, H. Wenz, H. Daerr, Th. Fritsch, Ch. Wellers, V. Korobov, D. Bakalov, S. Jorgensen, M. Okhapkin, and A. Nevsky. Technical support was provided by P. Dutkiewicz, R. Gusek, J. Bremer, and H. Hoffmann. We are grateful for financial support from the German Science Foundation, the EC network "Cold Molecules", the Alexander-von-Humboldt Foundation, the Düsseldorf Entrepreneurs Foundation, the DAAD, and the Studienstiftung des Deutschen Volkes.